\RequirePackage{float}
\documentclass[a4paper]{cas-sc}
\usepackage[english]{babel}
\addto\captionsenglish{}
\usepackage{csquotes}
\usepackage[backend=biber,style=apa,natbib=true,maxcitenames=1,uniquelist=false]{biblatex}
\DeclareSourcemap{
  \maps[datatype=bibtex]{
    \map{
      \pertype{misc}
      \step[fieldsource=archiveprefix, match=\regexp{arXiv}, final]
      \step[typesource=misc, typetarget=online]
      \step[fieldset=organization, fieldvalue={arXiv}]
    }
    \map{
      \step[fieldsource=entrykey, match=\regexp{\Almevalharness2024\z}, final]
      \step[typesource=misc, typetarget=software]
    }
  }
}
\AtEveryBibitem{\iffieldundef{doi}{}{\clearfield{eprint}\clearfield{url}\clearfield{urldate}}}
\usepackage{placeins}
\usepackage{adjustbox}
\usepackage{ragged2e}
\usepackage{needspace}
\usepackage{algorithm}
\usepackage{algpseudocode}
\algrenewcommand{\algorithmicrequire}{\textbf{Input:}}
\algrenewcommand{\algorithmicensure}{\textbf{Output:}}
\makeatletter
\AtBeginDocument{\@fleqnfalse\@mathmargin=\@centering}
\makeatother
\ifdefined\bibsep\else\newlength{\bibsep}\fi 
\AtBeginBibliography{\RaggedRight}
\graphicspath{{figures/}}

\newtheorem{proposition}{Proposition}
\providecommand{\Description}[1]{}
\newcommand{\rmt}{\rho_{\text{layer}}}
\newcommand{\sstar}{s^{\star}}
\newcommand{\papertablefont}{\footnotesize}
\usepackage{tikz}
\newcommand*\circledgreen[1]{\tikz[baseline=(char.base)]{
    \node[shape=circle, fill=ForestGreen, inner sep=1.3pt] (char) {\textcolor{white}{\textbf{#1}}};}}

\begin{document}
\let\WriteBookmarks\relax
\renewcommand{\floatpagefraction}{.80}
\renewcommand{\textfraction}{.05}
\renewcommand{\topfraction}{.92}
\renewcommand{\bottomfraction}{.85}
\setcounter{topnumber}{3}
\setcounter{bottomnumber}{2}
\setcounter{totalnumber}{5}
\makeatletter
\setlength{\@fptop}{0pt}
\setlength{\@fpbot}{0pt plus 1fil}
\makeatother
\setlength{\textfloatsep}{10pt plus 2pt minus 2pt}
\setlength{\floatsep}{10pt plus 2pt minus 2pt}
\setlength{\intextsep}{10pt plus 2pt minus 2pt}
\setlength{\tabcolsep}{5pt}
\renewcommand{\arraystretch}{0.94}

\shorttitle{AnchorKV}
\shortauthors{Li, Zhang, Qasim and H. Zhang}
\title[mode=title]{AnchorKV: Lossless Key-Value Cache Reconstruction from
Layer-Sparse Residual Anchors}

\ExplSyntaxOn
\RenewDocumentCommand \firstname {} { \color{black} \seq_use:Nn \l_stm_au_seq { ~ } }
\ExplSyntaxOff

\author[1]{Hao Li}[orcid=0009-0009-5907-9845]
\ead{hli@my.swjtu.edu.cn}
\author[1]{Jiashu Zhang}[orcid=0000-0003-2086-0991]
\cormark[1]
\ead{jszhang@home.swjtu.edu.cn}
\author[1]{Kaleem Ullah Qasim}[orcid=0000-0002-0102-3816]
\ead{kaleem@my.swjtu.edu.cn}
\author[1]{Heying Zhang}
\ead{hey_zhang@my.swjtu.edu.cn}
\cortext[1]{Corresponding author}
\address[1]{School of Computing and Artificial Intelligence, Southwest Jiaotong University, Chengdu, China}

\begin{abstract}
Residual-stream caching can reconstruct transformer key--value (KV) caches
losslessly, but its memory advantage depends on architecture. We present
AnchorKV, a storage--replay spectrum connecting per-layer residual caching and
full recomputation through configurable residual anchors. An
architecture-dependent break-even rule identifies the first storage-saving
spacing, while deployment selects an operating point from the spectrum and
cascade replay reduces reconstruction from $O(Ls)$ to $O(L)$,
with $L$ the model depth and $s$ the anchor spacing. For interleaved
sliding-window architectures, exact dynamic programming places anchors under
non-uniform reconstruction demand. Across CPU, CUDA, and Ascend NPU backends,
six architecture families, and fp32/bf16 arithmetic, evaluations verify
bit-exact KV reconstruction and unchanged continuation logits and tokens. At
$4$k context, cascade reconstruction is up to $16.3$ times faster than naive
replay; across eight applicable global-attention checkpoints at $32$k context,
deepest spacing yields $5.78$--$54.0$ times full-KV reactivatable session
capacity. Compared with seven external cache methods on a common model,
AnchorKV preserves full-cache retrieval and question-answering scores, whereas
lossy methods change output fidelity and task scores. The break-even spacing
is computable from a model's configuration file alone.
\end{abstract}

\begin{keywords}
Key-value cache \sep Residual stream \sep Transformer inference \sep Lossless reconstruction \sep Recomputation \sep Anchor placement
\end{keywords}

\hfuzz=120pt
\maketitle
\hfuzz=.1pt

\section{Introduction}

In long-context autoregressive transformer inference, the key--value (KV) cache
can become the dominant memory cost, and a large literature reduces it through eviction,
quantization, low-rank projection, or architectural change
\citep{h2o2023,kivi2024,palu2024,gqa2023}. A distinct, lossless line instead
\emph{reconstructs} KV on demand from a cheaper cached state. The strongest
recent instance proves that in a pre-norm transformer the keys and values at
every layer are deterministic projections of the residual stream, so caching
residual vectors and recomputing KV yields bit-identical inference
\citep{kvdirect2026}. System work reaches the same operating point from the
serving side, checkpointing per-layer activations and recomputing KV to save
memory or overlap I/O \citep{hybridserve2025,kvpr2024}.

The per-layer form of this idea, used by activation-checkpointing systems,
caches one residual state per layer and assumes that a layer's residual is
cheaper than its KV. That holds for multi-head attention, where a KV pair is twice the width of the
residual, but it fails for grouped-query attention. Our audit (\S\ref{sec:bound})
finds the residual-to-KV storage ratio spanning $0.50$--$3.50$ across deployed
models; Llama-3.1-8B, Mistral-7B, Gemma-3, and Qwen2.5 all require more residual
storage than KV. The shorthand ``cache residuals, save half the memory'' is
therefore an MHA special case, and on a modern $8$B model the same recipe doubles
memory.

This observation motivates a storage--replay spectrum: keep a residual
\emph{anchor} every $s$ layers, and rebuild a non-anchored layer by replaying
the transformer forward from the nearest preceding anchor. Larger spacing stores fewer anchors but requires
more replay. The endpoints recover known methods. At $s=1$, the scheme is
per-layer activation caching \citep{hybridserve2025}; at $s=L$, it is
single-vector residual checkpointing \citep{kvdirect2026}. GQA models first
fall below the KV baseline in the interior, which no prior method exposes and
which this paper formalizes and measures.

\begin{figure}
\centering
\includegraphics[width=\linewidth]{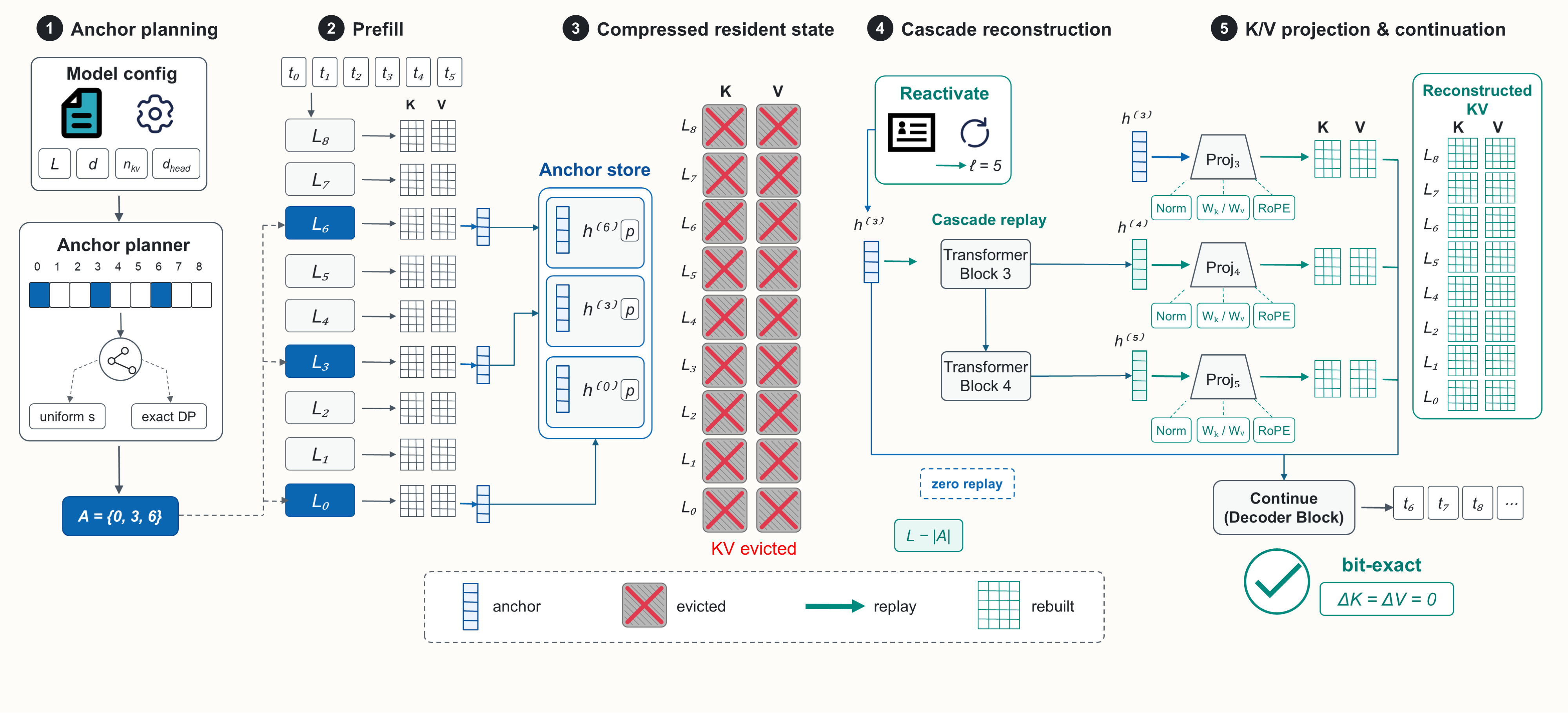}
\caption{\textbf{AnchorKV architecture.} A planner selects residual anchors;
prefill saves each anchor's input residual and position metadata, and the
prefix K/V is released. Reactivation projects K/V from each anchor and replays
the intervening blocks, restoring the cache for ordinary decoding.}
\Description{Five-stage left-to-right architecture diagram. First, model
configuration parameters feed an anchor planner that selects either uniform or
exact dynamic-programming anchors, illustrated by layers zero, three, and six.
Second, prefill computes layerwise keys and values while saving only the blue
anchor residuals and position metadata. Third, the layerwise keys and values
are crossed out to show eviction from the compressed resident state. Fourth,
reactivation restores the nearest preceding anchor and cascades through the
intervening transformer blocks. Fifth, normalization, key and value projection,
and rotary position encoding rebuild the cache before decoder continuation. A
check mark reports bit-exact reconstructed keys and values.}
\label{fig:overview}
\end{figure}

Figure~\ref{fig:overview} summarizes the end-to-end architecture, from anchor
planning and compressed residency to bit-exact reconstruction and continuation.

Our contributions are as follows.

\vspace{4pt}
\noindent\circledgreen{1}\;Per-layer residual caching increases memory on most
deployed models. Whether residuals beat KV is set by one config ratio,
$\rmt=d/(2n_{kv}d_{\text{head}})$, which spans \textbf{0.50--3.50} across six
audited families; the first strictly saving spacing is
$\sstar=\lfloor\rmt\rfloor+1$, and four measured crossings land exactly where
the rule predicts (\S\ref{sec:bound}).

\vspace{4pt}
\noindent\circledgreen{2}\;Residual anchors every $s$ layers span a lossless
storage--replay spectrum between per-layer caching ($s=1$) and KV-Direct
($s=L$), and cascade replay makes its deep end affordable: reconstruction falls
from $O(Ls)$ to $O(L)$ block executions ($276\to23$ on 24-layer Qwen2.5), and at
$4$k context cascade is \textbf{7.17--16.25$\times$} faster than naive replay at
$s=L$ across nine models (\S\ref{sec:replay}).

\vspace{4pt}
\noindent\circledgreen{3}\;Exact dynamic-programming placement serves the
non-uniform demand of interleaved sliding-window attention: on native
Gemma-3-1B it cuts replay from \textbf{18 layer executions to zero} and median
latency from $60.64$ to $5.52$\,ms, and predicted and measured executions agree
in all $24/24$ controlled comparisons (\S\ref{sec:placement}).

\vspace{4pt}
\noindent\circledgreen{4}\;Reconstruction is bit-exact
($\max|\Delta K|=\max|\Delta V|=0$) across \textbf{12 checkpoints} from six
model families across CPU, CUDA, and Ascend NPU in fp32 and bf16; all $441$
downstream traces are exact. At $32$k context the deepest spacing gives \textbf{5.78--54.0$\times$}
full-KV usable session capacity on eight models. Against seven external cache
methods on LongBench, AnchorKV keeps full-KV scores, while every lossy baseline
changes outputs, with $3.52$--$63.67\%$ argmax mismatch at $16$k
(\S\ref{sec:complete}).

\section{Research Objectives}\label{sec:objectives}

This study has four linked research objectives. First, we quantify when storing
residual anchors is cheaper than retaining full KV, and derive a
model-specific break-even spacing for each architecture. Second, we formalize the storage--replay spectrum and test whether
cascade replay makes the interior and deep-anchor operating points practical while
preserving bit-exact K/V, continuation logits, and tokens. Third, we determine how
anchors should be placed when reconstruction demand is non-uniform, as in
interleaved sliding-window attention, and verify the resulting dynamic-programming
solution on controlled and native Gemma-3 paths. Fourth, we measure the operational
consequence of the trade-off, including resident and reactivatable session capacity,
session switching, downstream equivalence, and comparison with lossy cache methods
under a common evaluation protocol. We evaluate these objectives with the
common protocols described in Section~\ref{sec:complete}. Recent EAAI survey work on
foundation-model systems likewise treats latency, memory/KV-cache use, and edge
scheduling as deployment constraints \citep{fourati2026foundation}.

\section{Related Work}\label{sec:related}

\citet{kvdirect2026} establish the per-layer projection identity we build on
and verify it bit-exactly across four architecture families. Their KV-Direct
scheme checkpoints one residual vector per evicted token, corresponding to
the $s=L$ residual-storage endpoint in our spectrum. The projection identity
applies whenever the target layer's residual is available, whether retained
or reconstructed; it underlies every spacing and does not define a separate
$s=1$ storage scheme. AnchorKV exposes intermediate spacings and makes their
storage and replay costs explicit. \citet{hybridserve2025} checkpoint per-layer activations
($s{=}1$ storage) in a vLLM-based serving stack and trade KV against activation
storage; \citet{kvpr2024} send partial activations over PCIe and recompute the
rest to overlap I/O. Both fix $s=1$ and neither replays across layers. \citet{slimattention2025} store only $K$ and recompute $V=K\,W_K^{-1}W_V$;
their residual-form saving is limited to cases where $d_{cache}>d_{model}$ and
can disappear as GQA narrows $n_{kv}$; this is the same dimensional boundary
captured by our residual-vs.-KV ratio across the whole spectrum. They also list
residual recomputation as incompatible with rotary position embeddings
\citep{rope2021}; our re-rotation mechanism removes that restriction
(\S\ref{sec:complete}). Token recomputation, measured by \citet{hybridserve2025}, occupies a similar
full-depth compute regime to $s=L$. Its latency motivates exploring intermediate
anchor spacings.

A separate study trains cacheless decoders by stabilizing residual streams,
rather than reconstructing an evicted KV cache at inference time; it is a
complementary cacheless design point \citep{qasim2026eaai}. Eviction \citep{h2o2023,streamingllm2023,tova2024,snapkv2024,adakv2025,pyramidkv2024},
quantization \citep{kivi2024,kvquant2024}, and low-rank projection
\citep{palu2024} reduce KV by altering retained information, with task-dependent
quality effects; we treat these as lossy
baselines orthogonal to and composable with anchoring. Architectural methods
shrink KV through training: MQA/GQA \citep{mqa2019,gqa2023},
cross-layer attention
\citep{cla2024}, layer-condensed KV \citep{lckv2024}, YOCO \citep{yoco2024}, and
MLA \citep{mla2024}, which can be read as a learned residual checkpoint. These
require retraining. MLA, for example, fixes its compression at pretraining
time and cannot be applied to the deployed GQA checkpoints in our audit,
whereas anchoring applies to any pre-norm checkpoint without retraining. Orthogonally, \citet{rmt2025}
enlarge the residual stream itself with an outer-product memory at training
time, whereas we exploit the existing residual stream at inference without
touching the architecture. \citet{pyramidkv2024} observe
that deep layers attend to fewer tokens, and \citet{minicache2024} that adjacent
deep layers have similar KV; we test the latter as a source of savings and find
it lossy (\S\ref{sec:replay}), which motivates exact replay. A separate line
learns compact substitutes for input context: gist tokens compress instruction
prompts, while AutoCompressors represent long contexts with short sequences of
soft prompts \citep{gist2023,autocompressors2023}. Their fidelity is evaluated through retained task performance. Anchoring preserves
each token's residual at every anchor layer, applies the projection identity to
exact inputs, and gives bit-exact fidelity by construction
($\max|\Delta K|=\max|\Delta V|=0$, \S\ref{sec:complete}). The residual
stream also linearly encodes belief-state geometry \citep{beliefstate2024};
our results are consistent with that finding but do not depend on it.

\section{Method}\label{sec:method}

\subsection{Storage--Replay Spectrum}\label{sec:spectrum}

Consider a pre-norm transformer with $L$ layers and a contiguous prefix of
$N$ tokens, starting at the beginning of the context. Token $t$ enters block
$l$ with residual $h_t^{(l)}\in\mathbb{R}^d$; write the entire prefix as
$H^{(l)}=[h_0^{(l)};\ldots;h_{N-1}^{(l)}]\in\mathbb{R}^{N\times d}$, omitting
the batch dimension. Choose anchor layers $A\subseteq\{0,\ldots,L-1\}$. For
a demand set $S$, the first anchor must satisfy $\min A\le\min S$; when $S$
includes layer 0, this reduces to $0\in A$. A saved anchor
$H_a=H^{(a)}$ is the \emph{input to block $a$, before its attention
normalization}. Thus every prefix token contributes one residual vector at
each selected layer. Figure~\ref{fig:overview} follows this state through
capture, eviction, reconstruction, and continuation.

During normal prefix evaluation, record $H_a$ whenever the forward pass enters
an anchor layer. Once the session becomes inactive, keep these matrices and
the position/attention metadata $P$, and release the prefix's per-layer K/V
and non-anchor intermediate activations. Model parameters remain available.
For the global-attention core, the retained session state is therefore
$\{H_a:a\in A\}$ plus $P$. The saving is across layer depth: all $N$ token
rows remain at every anchor. Local-window caches needed by hybrid architectures
are additional retained state, as specified in \S\ref{sec:gemma-serviceable}.

Given $H^{(l)}$, the original model modules directly reproduce layer $l$'s
cache \citep{kvdirect2026}:
\begin{equation}\label{eq:kv-input}
X_l = \operatorname{Norm}^{\mathrm{attn}}_l(H^{(l)}).
\end{equation}
\begin{equation}\label{eq:kv-from-residual}
K_l = \operatorname{PosK}_l\!\left(
  \operatorname{KNorm}_l(\operatorname{KProj}_l(X_l)),P\right).
\end{equation}
\begin{equation}\label{eq:v-from-residual}
V_l = \operatorname{VProj}_l(X_l).
\end{equation}
Here $\operatorname{KProj}_l(X)=XW_l^K+b_l^K$ and
$\operatorname{VProj}_l(X)=XW_l^V+b_l^V$ are the learned affine projections,
with the model's native head reshaping understood.
$\operatorname{KNorm}_l$ is the model's key normalization, or the identity
when absent; $\operatorname{PosK}_l$ applies its positional key transform,
such as RoPE at the original token positions. We denote this whole operation
by $\Pi_l(H^{(l)},P)=(K_l,V_l)$, called \textsc{ProjectKV} below.
At an anchor, $H^{(l)}=H_l$ is already saved, so obtaining this layer's K/V
requires only this projection path, with no preceding transformer block.

For a non-anchor layer $l$, start at its nearest preceding anchor $a<l$ and
evaluate the original blocks $F_a,\ldots,F_{l-1}$ on the entire prefix.
For the usual two-sublayer pre-norm block, the residual update is
\begin{equation}\label{eq:replay-query}
Q_j = \operatorname{Query}_j(X_j,P).
\end{equation}
\begin{equation}\label{eq:replay-attention}
U_j = H^{(j)}+\operatorname{Out}_j\!\left(
  \operatorname{Attn}_{P}(Q_j,K_j,V_j)\right).
\end{equation}
\begin{equation}\label{eq:replay-state}
H^{(j+1)} = U_j+\operatorname{MLP}_j\!\left(\operatorname{Norm}^{\mathrm{ffn}}_j(U_j)\right).
\end{equation}
The native query path includes query normalization and position transforms
where required; $\operatorname{Out}_j$ includes head concatenation and output
projection. Architecture-specific extra normalizations or residual scaling
remain part of the original $F_j$.
Replay does not read the evicted K/V: $H^{(j)}$ contains all prefix
tokens, so their layer-$j$ Q/K/V can be recomputed before causal attention is
evaluated. Every causal predecessor lies in this prefix. Attention and the MLP
then produce the next layer input, from which Eqs.~\eqref{eq:kv-from-residual}--\eqref{eq:v-from-residual}
recover the next cache. Queries, attention outputs, and non-anchor residuals
are temporary replay state. They need not be retained while the session is cold.

Reconstruction emits $\widehat{\mathcal C}=\{(K_l,V_l)\}_{l=0}^{L-1}$ in the
cache's original token order. The next input token is processed at position
$N$ using this recovered history, and its new K/V entries are appended through
ordinary autoregressive decoding. No vocabulary projection of the old prefix
is needed to restore its cache. Once the session is active, subsequent
tokens use that cache; reconstruction is triggered by reactivation after
eviction, not by every generated token. Section~\ref{sec:replay} explains how
to share replay among all missing layers and states the numerical conditions.

With uniform spacing $s$, the per-token storage and worst-case replay depth are
\begin{equation}\label{eq:storage-bytes}
M(s)=\lceil L/s\rceil\cdot d\cdot b \quad\text{bytes/token}.
\end{equation}
\begin{equation}\label{eq:replay-depth}
R(s)\le (s-1)\ \text{forward layers} + 1\ \text{projection}.
\end{equation}
In Eqs.~\eqref{eq:storage-bytes}--\eqref{eq:replay-depth}, $b$ is bytes per element. Relative to a standard KV cache of
$2Ln_{kv}d_{\text{head}}b$ bytes/token, the exact storage ratio is
$\lceil L/s\rceil\rmt/L$ and the all-layer cascade replay fraction is
$1-\lceil L/s\rceil/L$; these reduce to $\rmt/s$ and $1-1/s$ when $s\mid L$.
Table~\ref{tab:endpoints} maps the endpoints onto prior work.

Two accounting rules hold throughout. First, every storage ratio is taken
against bytes the baseline actually keeps resident for the same tokens on the
same device. The baseline is the full per-layer KV cache above, and the capacity
results of \S\ref{sec:capacity} are measured to the out-of-memory boundary. Second, nothing the deployed model already declines to
store is counted as savings: in interleaved sliding-window architectures the
window layers retain only a bounded window of KV, so beyond that window the
comparison restricts to the global layers that do retain it. These global
layers form the non-uniform demand set treated in \S\ref{sec:replay}.

\begin{table}
\centering\papertablefont
\caption{Uniform residual-anchor configurations. Replay depth counts
transformer blocks from an anchor at layer 0, excluding the K/V projection.}
\label{tab:endpoints}
\maxsizebox{\linewidth}{!}{%
\begin{tabular}{lccp{15.5pc}}
\toprule
$s$ & residual storage/token & max. replay depth & storage scheme \\
\midrule
$s=1$ & $L\,d\,b$ & $0$ & Per-layer activation cache \citep{hybridserve2025} \\
$1<s<L$ & $\lceil L/s\rceil\,d\,b$ & $\le s-1$ & \textbf{Intermediate residual anchors (AnchorKV)} \\
$s=L$ & $d\,b$ & $L-1$ & KV-Direct \citep{kvdirect2026} \\
\bottomrule
\end{tabular}
}
\end{table}

\begin{figure}
\centering
\includegraphics[width=\linewidth]{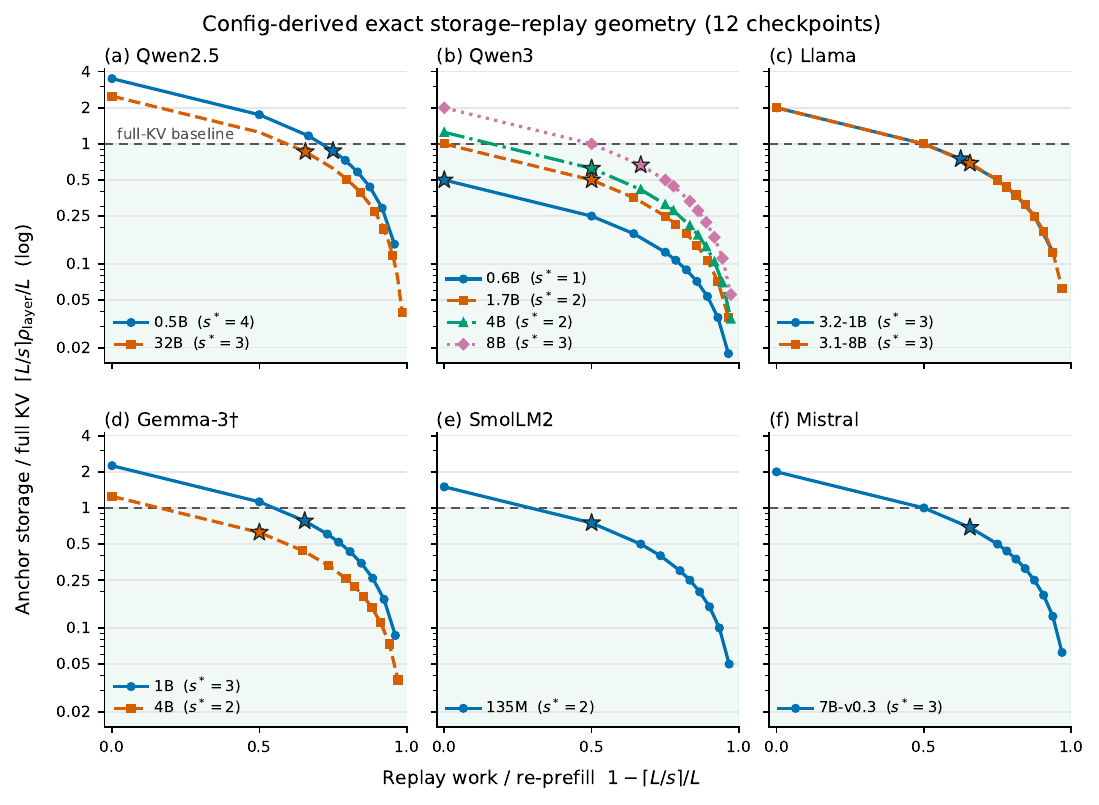}
\caption{\textbf{Storage--replay geometry for 12 checkpoints in six model families.}
Curves plot normalized replay $1-A(s)/L$ against storage $A(s)\rmt/L$ (log
axis), with $A(s)=\lceil L/s\rceil$. Stars mark $\sstar$, the first spacing
below the dashed full-KV line. $\dagger$ Gemma-3 uses a dense-equivalent
baseline (native results: Table~\ref{tab:gemma-longcontext}).}
\Description{Six-panel small-multiple line chart of exact discrete storage and
replay geometry for twelve checkpoints grouped into Qwen2.5, Qwen3, Llama,
Gemma-3, SmolLM2, and Mistral families. A star on each model curve marks the
first integer spacing strictly below the dashed full-KV baseline.}
\label{fig:spectrum}
\end{figure}

Figure~\ref{fig:spectrum} groups every checkpoint by family and is computed
from configurations alone; measured performance appears in \S\ref{sec:complete}.
The detailed all-spacing evaluation (Qwen3-0.6B, SmolLM2-135M,
Qwen3-8B, and Qwen2.5-0.5B) establishes that the interior is traversable and
bit-exact; at $s{=}L$, the independent per-layer reconstruction in the 24-layer
Qwen2.5 verification accumulates $276$ block executions. Sharing replay reduces
that count to $23$ (\S\ref{sec:replay}). A separate Mistral-7B-v0.3 reconstruction experiment on Ascend 910B4
(bf16, SDPA) verifies reconstruction: $32/32$ layer K/V checks, $30/30$
continuation-logit checks, and $30/30$ greedy-token checks are exact, while all $56$
timing rows over 256--4096-token contexts are lossless (all $32$ spacings at
256 tokens, plus six standard
spacings at each longer context). This experiment is separate from the
nine-model evaluation, which measures Mistral under the shared protocols. Native Gemma residency and serving use
a separate protocol that retains sliding-window KV and position state and derives
a native resident break-even spacing.

\subsection{Storage Criterion and Break-Even Spacing}\label{sec:bound}

\begin{proposition}[Net storage benefit]\label{prop:bound}
Anchoring at uniform spacing $s$ with $s\mid L$ saves memory over the standard KV
cache iff
\begin{equation}
\label{eq:storage-benefit}
\frac{L}{s}\,d\,b < 2\,L\,n_{kv}\,d_{\text{head}}\,b
\quad\Longleftrightarrow\quad
s > \rmt \triangleq \frac{d}{2\,n_{kv}\,d_{\text{head}}}.
\end{equation}
For general $s$ the anchor count is $\lceil L/s\rceil$ and the condition is
$\lceil L/s\rceil\,\rmt < L$, which implies $s>\rmt$.
\end{proposition}
Equation~\eqref{eq:storage-benefit} follows directly by comparing the storage
in Eq.~\eqref{eq:storage-bytes} with full KV. Because the inequality is
strict, the smallest integer spacing that pays under the $s\mid L$ form is
$\sstar=\lfloor\rmt\rfloor{+}1$, the least integer \emph{exceeding} $\rmt$: at
integer $\rmt$ the spacing $s=\rmt$ only ties the KV cache
(Figure~\ref{fig:spectrum} shows Qwen3-8B, $\rmt{=}2.00$, at exactly
$1.00\times$ for $s{=}2$), and $\rmt<1$ gives $\sstar=1$. For general $s$ the
ceiling can defer the first strict saving at small $L$, so the operative
per-model check is $\lceil L/s\rceil\,\rmt<L$. This still requires only one
line from a config file; every $\sstar$ in Table~\ref{tab:audit} passes it.

Because $\rmt$ varies across deployed architectures, no fixed $s$ breaks even
for all models; the break-even spacing $\sstar$ must be computed per model.

The break-even spacing is a storage threshold. For uniform global attention,
a deployment can choose $\sstar$, an interior
spacing, or $s=L$ according to its available memory, reactivation-latency
budget, session workload, and required reconstruction demand.

Table~\ref{tab:audit} reports $\rmt$ from published configs (no weights needed).
It spans $7\times$; the single model with $\rmt<1$ is Qwen3-0.6B, whose unusually
wide KV makes $s=1$ save half the memory: the MHA-like regime in which the
residual-caching intuition holds. Every other entry in
Table~\ref{tab:audit}, spanning the deployed Llama-3.1, Mistral, Gemma-3, and
Qwen2.5 families, has
$\rmt>1$, so per-layer residual caching costs them memory. As a
cross-check, our audit reproduces the Gemma-3 KV-per-token figure reported by
\citet{kvdirect2026} to the byte.

\begin{table}
\centering\papertablefont
\caption{Storage ratio $\rmt$ and break-even spacing $\sstar$ from model
configs (bf16).}
\label{tab:audit}
\maxsizebox{\linewidth}{!}{%
\begin{tabular}{lcccccc}
\toprule
model & $L$ & $n_{kv}$ & KV/tok & residual($s{=}1$)/tok & $\rmt$ & $\sstar$ \\
\midrule
Qwen3-0.6B & 28 & 8 & 112\,KB & 56\,KB & 0.50 & 1 \\
Gemma-3-4B & 34 & 4 & 136\,KB & 170\,KB & 1.25 & 2 \\
SmolLM2-135M & 30 & 3 & 22.5\,KB & 33.8\,KB & 1.50 & 2 \\
Llama-3.1-8B / Mistral-7B & 32 & 8 & 128\,KB & 256\,KB & 2.00 & 3 \\
Qwen2.5-32B & 64 & 8 & 256\,KB & 640\,KB & 2.50 & 3 \\
Qwen2.5-0.5B & 24 & 2 & 12\,KB & 42\,KB & 3.50 & 4 \\
\bottomrule
\end{tabular}
}
\par\smallskip
\begin{minipage}{\linewidth}
\papertablefont\raggedright
\textit{Configuration sources:}
\citep{qwen3_2025,gemma3_2025,smollm2_2025,llama3_2024,mistral7b_2023,qwen25_2024}.
\end{minipage}
\end{table}

Proposition~\ref{prop:bound} is an identity that converts a config file into the
break-even operating point without touching weights. Every star in
Figure~\ref{fig:spectrum} is therefore the first exact discrete point below the
KV baseline. The measured all-spacing sweeps independently validate four crossings:
Qwen3-0.6B at $s{=}1$ ($0.50\times$), SmolLM2-135M at $s{=}2$ ($0.75\times$),
Qwen3-8B ($\rmt{=}2.00$) at $s{=}3$ after exactly $1.00\times$ at $s{=}2$, and
Qwen2.5-0.5B at $s{=}4$ ($0.88\times$). The empirical content lies in the
audit's range: deployed $\rmt$ reaches $3.5$, so the regime where $s{=}1$
suffices ($\rmt<1$) is the exception, and every other audited model needs an
interior spacing to save memory.

\subsection{Cascade Replay}\label{sec:replay}

\subsubsection{Cascade replay: \texorpdfstring{$O(L)$}{O(L)} layer executions}
Reconstructing every evicted layer independently, each from its nearest anchor,
costs $O(L\!\cdot\!s/2)$ layer executions and grows with $s$, which would make
the deep, high-compression end of the spectrum impractical. Instead, partition
layer depth into segments $[a_i,a_{i+1})$ between successive anchors, with
$a_k=L$. Within one segment, load $H_{a_i}$ once, emit K/V from the current
residual using Eqs.~\eqref{eq:kv-from-residual}--\eqref{eq:v-from-residual}, and execute the next block only
if another layer input in that segment is needed. After emitting the last
layer's K/V, load the next saved anchor directly. The block connecting to that
anchor need not run: its output has already been saved. At the final segment,
the last block's output is also unnecessary for reconstructing K/V.

For an eight-layer example, let $L=8$, $s=4$, and $A=\{0,4\}$. Cold storage contains only $H_0$, $H_4$,
and $P$. The two replay paths are shown below; boxes mark saved inputs,
horizontal arrows are complete blocks, and downward arrows emit K/V.
\[
\begin{array}{ccccccc}
\boxed{H_0}&\xrightarrow{F_0}&H^{(1)}&\xrightarrow{F_1}&H^{(2)}&\xrightarrow{F_2}&H^{(3)}\\
\downarrow\Pi_0&&\downarrow\Pi_1&&\downarrow\Pi_2&&\downarrow\Pi_3\\
(K_0,V_0)&&(K_1,V_1)&&(K_2,V_2)&&(K_3,V_3)
\end{array}
\]
\[
\begin{array}{ccccccc}
\boxed{H_4}&\xrightarrow{F_4}&H^{(5)}&\xrightarrow{F_5}&H^{(6)}&\xrightarrow{F_6}&H^{(7)}\\
\downarrow\Pi_4&&\downarrow\Pi_5&&\downarrow\Pi_6&&\downarrow\Pi_7\\
(K_4,V_4)&&(K_5,V_5)&&(K_6,V_6)&&(K_7,V_7)
\end{array}
\]
This restores all eight layers' caches using six complete blocks:
$F_0,F_1,F_2,F_4,F_5,F_6$. Skipping $F_3$ and $F_7$ does not omit their K/V:
$\Pi_3$ and $\Pi_7$ already emit those entries. Independently replaying from
each anchor for each target would instead execute $0+1+2+3$ blocks per
segment, or twelve blocks in total. Cascade shares those intermediate residuals.

Algorithm~\ref{alg:cascade-replay} gives the global-attention reconstruction
procedure. \textsc{ProjectKV} is given by Eqs.~\eqref{eq:kv-from-residual}--\eqref{eq:v-from-residual};
\textsc{ReplayBlock} evaluates the original block on the prefix as in
Eqs.~\eqref{eq:replay-query}--\eqref{eq:replay-state}, including causal attention and the MLP.
At every projection, $H$ is the input to the target layer, not its output.
Each segment executes $a_{i+1}-a_i-1$ complete blocks, giving $E=L-|A|$.
This count excludes the separate K/V projections. The pseudocode separates
cache emission from residual propagation for clarity: a native block call may
recompute projections already emitted by \textsc{ProjectKV}. The bound counts
complete block executions; fused-kernel and wall-clock performance depend on
the implementation.

\begin{algorithm}
\caption{Cascade KV reconstruction for global attention}
\label{alg:cascade-replay}
\small
\begin{algorithmic}[1]
\Require Native blocks $F_0,\ldots,F_{L-1}$; demand set $S$; anchors
$0\le a_0<\cdots<a_{k-1}<L$ with $a_0\le\min S$
\Require Saved full-prefix input matrices $\{H_{a_i}\}$; position/attention metadata $P$
\Ensure Reconstructed K/V cache $\widehat{\mathcal C}$; block-execution count $E$
\State $a_k\gets L$; $\widehat{\mathcal C}\gets\emptyset$; $E\gets0$
\For{$i=0,\ldots,k-1$}
  \State $H\gets H_{a_i}$
  \For{$l=a_i,\ldots,a_{i+1}-1$}
    \State $\widehat{\mathcal C}_{l}\gets$ \Call{ProjectKV}{$l,H,P$}
    \If{$l<a_{i+1}-1$}
      \State $H\gets$ \Call{ReplayBlock}{$F_l,H,P$}
      \State $E\gets E+1$
    \EndIf
  \EndFor
\EndFor
\State \Return $\widehat{\mathcal C},E$
\end{algorithmic}
\end{algorithm}

\paragraph{Execution conditions and hybrid attention.}
Full-prefix replay supplies all token rows at once. If tokens are processed in
chunks, replay them chronologically and keep the \emph{newly reconstructed}
K/V of earlier chunks in a temporary cache, so later chunks see the correct
causal history. Neither execution needs a copy of the evicted original K/V.
Bit-exact restoration additionally requires unchanged weights and matched
precision, kernels, positions, attention masks, and batching/chunking.
In particular, repacking a history produced by tokenwise decoding into a
different full-prefix replay schedule can change floating-point results;
repeated eviction of an extended history requires matching that schedule or
separate validation. The measured exactness results apply to the execution
conditions tested in \S\ref{sec:complete}.

For a demand subset $S$, skip segments without targets, stop each remaining
segment at its last demanded layer, and emit K/V only for layers in $S$.
Native Gemma reconstruction additionally retains local-window KV and follows
its native local/global attention and position rules. The algorithm above gives
the global-attention core; the native path combines it with the local-window
cache and position state.

\subsection{Anchor Placement}\label{sec:placement}
Given a demand set $S$ of layers that must be reconstructed and a per-token
anchor budget $B$, we place anchors to minimize cascade replay. An anchor set
$A$ partitions the depth into segments $[a,t)$, ending at the next anchor or
$L$. The subset form of Algorithm~\ref{alg:cascade-replay} traverses each
segment only to its farthest demanded layer. For nonempty $S$, a feasible
first anchor must satisfy $\min A\le\min S$, and the objective is
\begin{equation}\label{eq:anchor-objective}
\begin{alignedat}{2}
C(a,t)&=\max\!\left(\{l-a:l\in S\cap[a,t)\}\cup\{0\}\right),\\
\min_A\quad &\sum_{[a,t)\in\mathrm{seg}(A)}C(a,t)
&&\text{s.t.}\quad |A|\,d\,b\le B,\quad \min A\le \min S.
\end{alignedat}
\end{equation}
We use $k=|A|$ for the explicit number of stored anchors. When a policy is
specified by a uniform spacing $s$, its anchor set is
$A_s=\{0,s,2s,\ldots\}\cap[0,L)$ and its count is
$k_s=\lceil L/s\rceil$; $s$ and $k$ are therefore different policy
parameters. DP optimizes over arbitrary anchor sets at a fixed $k$. For a
non-uniform demand set, $k=|S|$ permits $A=S$ and hence zero cascade replay,
whereas uniform spacing generally does not place every anchor on a demanded
layer. Capacity comparisons at the same $k$ use the same number of residual
states; placement changes replay work, not resident-state bytes.
Here $b$ retains its meaning of bytes per element, whereas $t$ denotes a
segment boundary. Let $k=\min\{L,\lfloor B/(db)\rfloor\}$. If $S$ is empty,
zero anchors and zero replay suffice; if $S$ is nonempty and $k=0$, the budget
is infeasible. Otherwise, adding anchors cannot increase the cost in
Eq.~\eqref{eq:anchor-objective}, so an optimum can use exactly $k$ anchors.

Define $D_i(a)$ as the minimum cost from layer $a$ to $L$ using $i$ anchors,
with the first at $a$. The recurrence and optimum are
\begin{equation}\label{eq:placement-dp}
\begin{alignedat}{2}
D_1(a)&=C(a,L),
&&D_i(a)=\min_{a<t<L}\!\bigl\{C(a,t)+D_{i-1}(t)\bigr\},\quad 2\le i\le k,\\
D^\star&=\min_{0\le a\le \min S}D_k(a).
\end{alignedat}
\end{equation}

Algorithm~\ref{alg:anchor-placement} evaluates this recurrence and returns both
the optimal first anchor and the complete anchor set recovered from its successor
pointers.

\begin{algorithm}
\caption{Exact dynamic-programming anchor placement}
\label{alg:anchor-placement}
\small
\begin{algorithmic}[1]
\Require Model depth $L$; demand set $S$; per-token budget $B$; residual width $d$; bytes per element $b$
\Ensure Optimal anchor set $A^\star$ and minimum cascade cost $D^\star$
\State $k\gets\min\{L,\lfloor B/(db)\rfloor\}$
\If{$S=\emptyset$}
  \State \Return $A^\star\gets\emptyset$, $D^\star\gets0$
\EndIf
\If{$k=0$}
  \State \Return \textsc{Infeasible}
\EndIf
\For{$a=0,\ldots,L-1$}
  \For{$t=a+1,\ldots,L$}
    \State $C(a,t)\gets\max\!\left(\{l-a:l\in S\cap[a,t)\}\cup\{0\}\right)$
  \EndFor
\EndFor
\For{$a=0,\ldots,L-1$}
  \State $D_1(a)\gets C(a,L)$
\EndFor
\For{$i=2,\ldots,k$}
  \For{$a=0,\ldots,L-i$}
    \State $\pi_i(a)\gets\arg\min_{a<t\le L-i+1}\bigl\{C(a,t)+D_{i-1}(t)\bigr\}$
    \State $D_i(a)\gets C\bigl(a,\pi_i(a)\bigr)+D_{i-1}\bigl(\pi_i(a)\bigr)$
  \EndFor
\EndFor
\State $a_0^\star\gets\arg\min_{0\le a\le\min S}D_k(a)$; $a\gets a_0^\star$
\State $A^\star\gets\{a\}$
\For{$i=k,k-1,\ldots,2$}
  \State $a\gets\pi_i(a)$; $A^\star\gets A^\star\cup\{a\}$
\EndFor
\State $D^\star\gets D_k(a_0^\star)$
\State \Return $A^\star,D^\star$
\end{algorithmic}
\end{algorithm}

The loop bounds exclude invalid states (equivalently, they have value
$+\infty$); ties in an $\arg\min$ may be resolved deterministically. The
pointer $\pi_i(a)$ is the next anchor after a first anchor at $a$, so the final
loop reconstructs the complete ordered anchor set; when $k=1$, that loop is
empty and $A^\star=\{a_0^\star\}$. Precomputing $C(a,t)$ in $O(L^2)$ time gives
an $O(kL^2)$ DP with $O(kL)$ state/pointer space, or $O(L^2+kL)$ total space
including the cost table. This bound assumes constant-time segment-cost lookup;
the direct reference implementation rescans the demand set when evaluating
segment costs.

For independent per-layer requests with frequencies $w_l\ge0$ and
$S=\{l:w_l>0\}$, replacing
$C(a,t)$ by $\sum_{l=a}^{t-1}w_l(l-a)$ gives a weighted objective with the same
recurrence. This counts independent replay work rather than one shared
cascade. All placement results below optimize and measure the cascade
objective in Eq.~\eqref{eq:anchor-objective}.

\begin{proposition}[Placement invariance]\label{prop:p1}
Under uniform demand ($S$ = all layers), the cascade replay
cost is $L-|A|$ layer executions for any anchor set of size $|A|$, independent of
placement: each segment walks to its last layer, and the walks tile the depth.
\end{proposition}
We verify this directly: four placements (uniform, shifted, front-loaded, random)
give identical execution counts, latencies within $11\%$, and bit-exact output. So
when demand is uniform, uniform placement is already optimal, and the DP solution
attains the same optimum cost. Placement matters only when demand is not uniform.

\begin{proposition}[DP benefit under non-uniform demand]\label{prop:p2}
When only a subset $S$ of layers must be reconstructed, exact DP placement reaches
zero replay once the budget $k=|S|$, by placing anchors at the layers in $S$.
\end{proposition}

\begin{figure}
\centering
\includegraphics[width=\linewidth]{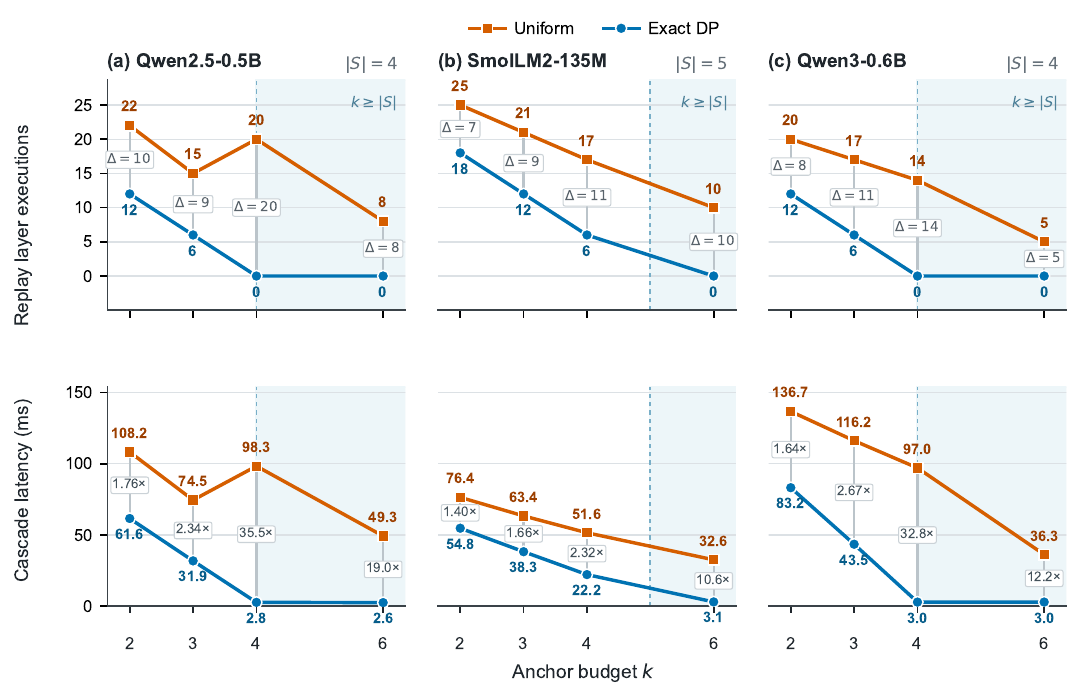}
\caption{\textbf{Anchor placement under Gemma-3-style demand}, transplanted to
three models at matched budgets. Top: layer executions, labelled with
uniform$-$DP. Bottom: cascade latency (ms), labelled with the uniform/DP ratio.
Shading marks $k\geq|S|$.}
\Description{Three paired point-line panels compare measured replay executions
under uniform and dynamic-programming placement for Qwen2.5, SmolLM2, and Qwen3.
Upper-row vertical connectors expose the per-budget reduction and carry the
uniform-minus-DP execution gap; each point is numerically labeled. Lower-row
panels plot the absolute measured cascade latencies in milliseconds, with
connector labels reporting the Uniform-to-Exact-DP latency ratio. Shaded
regions mark $k\geq|S|$. Exact DP requires fewer executions in every panel and
reaches zero at sufficient budget.}
\label{fig:dp}
\end{figure}

Interleaved sliding-window attention produces this kind of demand. In Gemma-3, one layer in six uses global attention and
the rest use local windows whose KV lives entirely inside the window and is never
part of eviction-driven reconstruction; the config marks the global layers as
$\{5,11,17,23,29\}$. The needed set is therefore the global layers.
Figure~\ref{fig:dp} applies this controlled transplant to Qwen2.5-0.5B,
SmolLM2-135M, and Qwen3-0.6B. At matched budget, DP placement uses fewer
executions than uniform and reaches zero replay once the budget covers the
needed set. Predicted and measured execution counts agree on all $24$
comparisons ($4$ budgets $\times$ uniform/DP $\times$ $3$ models). At
$k{=}|S|{=}4$ on Qwen2.5-0.5B, for example, uniform placement still needs $20$
layer executions where DP needs none, a measured $35.5\times$
cascade-latency reduction. Interleaved architectures are therefore both where
residual-only $s=1$ reconstruction fails (\S\ref{sec:complete}) and where
placement reduces replay the most.

The controlled transplant in Figure~\ref{fig:dp} is complemented by a direct
native Gemma-3 placement check. For Gemma-3-1B, the global-demand set is
$S_{1B}=\{5,11,17,23\}$. With budget $k=4$, evenly spaced uniform anchors
$\{0,6,13,19\}$ require $18$ predicted and measured layer executions, with a
median cascade latency of $60.64$\,ms. Exact DP places anchors at
$\{5,11,17,23\}$, reducing the predicted and measured execution count to zero
and the median latency to $5.52$\,ms. For Gemma-3-4B,
$S_{4B}=\{5,11,17,23,29\}$; uniform placement requires $19$ executions and
$70.59$\,ms, whereas native DP places anchors at the five global layers and
requires zero replay with a median latency of $7.41$\,ms. All four native
placement settings reproduce K/V and continuation outputs exactly. Thus,
Figure~\ref{fig:dp} isolates the placement optimizer on controlled dense
transplants, while these measurements verify the same placement rule inside
the native interleaved Gemma-3 cache.

\section{Experiments}\label{sec:complete}

\subsection{Experimental Setup and Protocol Coverage}

Table~\ref{tab:modelwide-coverage} covers 12 checkpoints and nine evaluation
protocols. Of the $108$ model--protocol combinations, $94$ provide validated
results. One direct-reuse control is inapplicable because Qwen3-0.6B anchors
every layer at $s^\star=1$. Four evaluations require contexts beyond SmolLM2's
limit: capacity, serving, long-context timing, and endpoint repetitions.
The remaining nine combinations, all for Qwen2.5-32B, exceed the
available memory of the 32\,GB Ascend device. Qwen2.5-32B is evaluated
separately on an A100 80\,GB for a scale audit. The Gemma evaluation includes
$18$ short-context measurements, $14$ long-context/residency measurements, and
$20$ serving and repeated measurements. Calibration and preliminary resource
probes set protocol parameters and serviceable loads; performance means use the
measurement runs.

\begin{table}
\centering\papertablefont
\caption{Coverage across 12 checkpoints: R=reconstruction; T=256--4096 timing;
C=32k capacity; S=32k serving; L=8k--32k timing; D=downstream equivalence;
A=cascade/naive control; N=direct-reuse control; E=endpoint repetitions.
\textsc{N/A-A}/\textsc{N/A-C}: inapplicable by architecture/context;
\textsc{RC}: exceeds device memory.}
\label{tab:modelwide-coverage}
\setlength{\tabcolsep}{3.5pt}
\maxsizebox{\linewidth}{!}{%
\begin{tabular}{@{}l*{9}{c}@{}}
\toprule
model & Recon. & Timing & \shortstack{32K\\capacity} & \shortstack{32K\\serving} & \shortstack{Long\\context} & Downstream & \shortstack{Cascade\\control} & \shortstack{Reuse\\control} & Repeats \\
\midrule
Qwen2.5-0.5B & \textsc{OK} & \textsc{OK} & \textsc{OK} & \textsc{OK} & \textsc{OK} & \textsc{OK} & \textsc{OK} & \textsc{OK} & \textsc{OK} \\
Qwen2.5-32B & \textsc{RC-HW} & \textsc{RC-HW} & \textsc{RC-HW} & \textsc{RC-HW} & \textsc{RC-HW} & \textsc{RC-HW} & \textsc{RC-HW} & \textsc{RC-HW} & \textsc{RC-HW} \\
Qwen3-0.6B & \textsc{OK} & \textsc{OK} & \textsc{OK} & \textsc{OK} & \textsc{OK} & \textsc{OK} & \textsc{OK} & \textsc{N/A-A} & \textsc{OK} \\
Qwen3-1.7B & \textsc{OK} & \textsc{OK} & \textsc{OK} & \textsc{OK} & \textsc{OK} & \textsc{OK} & \textsc{OK} & \textsc{OK} & \textsc{OK} \\
Qwen3-4B & \textsc{OK} & \textsc{OK} & \textsc{OK} & \textsc{OK} & \textsc{OK} & \textsc{OK} & \textsc{OK} & \textsc{OK} & \textsc{OK} \\
Qwen3-8B & \textsc{OK} & \textsc{OK} & \textsc{OK} & \textsc{OK} & \textsc{OK} & \textsc{OK} & \textsc{OK} & \textsc{OK} & \textsc{OK} \\
Llama-3.2-1B-Instruct & \textsc{OK} & \textsc{OK} & \textsc{OK} & \textsc{OK} & \textsc{OK} & \textsc{OK} & \textsc{OK} & \textsc{OK} & \textsc{OK} \\
Llama-3.1-8B-Instruct & \textsc{OK} & \textsc{OK} & \textsc{OK} & \textsc{OK} & \textsc{OK} & \textsc{OK} & \textsc{OK} & \textsc{OK} & \textsc{OK} \\
Gemma-3-1B & \textsc{OK} & \textsc{OK} & \textsc{OK} & \textsc{OK} & \textsc{OK} & \textsc{OK} & \textsc{OK} & \textsc{OK} & \textsc{OK} \\
Gemma-3-4B & \textsc{OK} & \textsc{OK} & \textsc{OK} & \textsc{OK} & \textsc{OK} & \textsc{OK} & \textsc{OK} & \textsc{OK} & \textsc{OK} \\
SmolLM2-135M & \textsc{OK} & \textsc{OK} & \textsc{N/A-C} & \textsc{N/A-C} & \textsc{N/A-C} & \textsc{OK} & \textsc{OK} & \textsc{OK} & \textsc{N/A-C} \\
Mistral-7B-v0.3 & \textsc{OK} & \textsc{OK} & \textsc{OK} & \textsc{OK} & \textsc{OK} & \textsc{OK} & \textsc{OK} & \textsc{OK} & \textsc{OK} \\
\bottomrule
\end{tabular}
}
\end{table}

\subsection{Datasets and Evaluation Inputs}\label{sec:datasets}

This is a model- and benchmark-based study; no new human participants or private
corpus is collected. The evaluation inputs are fixed before measurement and fall
into three groups. The model checkpoint corpus contains twelve public checkpoints
from six families: nine global-attention checkpoints (Qwen2.5-0.5B, Qwen3-0.6B,
Qwen3-1.7B, Qwen3-4B, Qwen3-8B, Llama-3.2-1B-Instruct,
Llama-3.1-8B-Instruct, SmolLM2-135M, and Mistral-7B-v0.3), two native
interleaved Gemma-3 checkpoints (1B and 4B), and Qwen2.5-32B for the scale
audit. The common nine-model panel supplies the reconstruction, timing,
capacity, serving, repeatability, and downstream protocols; Gemma-3 supplies
native sliding-window validation; Qwen2.5-32B is used only for the separately
identified A100 scale audit. All evaluations use fixed public checkpoint revisions
and fixed configuration settings; no checkpoint or runtime component is selected
after observing a method's output.

The downstream inputs are public benchmark examples. The cross-model equivalence
protocol fixes 49 traces per checkpoint across HellaSwag, PIQA, ARC-Easy,
ARC-Challenge, WinoGrande, LAMBADA, and WikiText long-generation traces. This
gives $9\times49=441$ fixed traces: 72 traces for each of the six standard
benchmarks and nine WikiText traces. The representative full-cache quality table
uses the same six standard benchmarks in zero-shot mode on Qwen2.5-0.5B. The
additional single-model audit contains 1765 benchmark examples and 121
32-step greedy generations; these are evaluation inputs, not training data.

The external cache-method comparison uses the public LongBench collection on
Qwen3-1.7B: English passage retrieval, Qasper, and HotpotQA, with 180 fixed
formal examples per task and full prompts retained. Retrieval prompts fit within
16k tokens; the formal Qasper and HotpotQA inputs span 2,087--20,210 and
1,910--17,652 tokens. A separate WikiText panel uses 16 fixed targets at both
4k and 16k prefixes with 32 teacher-forced steps. All benchmark scores use the
official task scorers; no example is selected after observing a method's output.

\paragraph{Cross-model validation protocol.}
We use fixed public revisions of the nine global-attention checkpoints and a
common runtime configuration;
Table~\ref{tab:modelwide-coverage} summarizes measured results and architecture, context, and memory limits. We run
the nine-model panel in bf16 with Hugging Face SDPA on the same
$1\times$ Ascend 910B4 32\,GB NPU. The short-context sweep uses
$N\in\{256,512,1024,2048,4096\}$ and the de-duplicated spacing set
$\{1,2,4,8,\sstar,L\}$. Its $45$ independent model--context runs contain
$245$ operating-point rows. Every latency is the median of five synchronized
repetitions after one warmup; anchors and reference KV come from the same
cache-producing prefill, and K/V, continuation logits, and greedy continuation
tokens are checked bitwise.

The long-context experiments use a separate measurement implementation at
$N\in\{8192,16384,32768\}$ and $s\in\{\sstar,8,L\}$: eight applicable models,
$24$ runs in fresh processes, and $72$ rows. SmolLM2 is \textsc{N/A--Context}
because its 8192-token limit cannot hold the required continuation. Both the
short- and long-context protocols disable autograd. The short-context protocol
completes its exactness checks before timing; the long-context protocol moves
reference K/V and logits to host memory and releases their device copies before
replay. Setup and reference transfer are excluded from replay timing. Because
the two protocols differ in measurement implementation and tensor lifetime, we
report their process-level memory peaks separately
(Table~\ref{tab:crossmodel-peak-memory}). The capacity
protocol likewise covers all eight applicable
models under full KV and $s\in\{\sstar,8,L\}$, yielding $32$ measured points.
Each model--context or capacity configuration is measured in a separate
process. Reported results pass independent checks of parameter settings,
measurement completeness, and reconstruction exactness.

\subsection{Bit-Exact Reconstruction}

\paragraph{Position encoding $\times$ attention completeness.}
Reconstruction from a residual anchor is exact for global attention with RoPE
because the projection and rotation are deterministic. Sliding-window layers
break the residual-only recipe twice: they rotate with their own base
($\theta{=}10^4$ locally vs.\ $10^6$ globally in Gemma-3), and the position
index their rotation consumed is not recoverable from the residual once the
window rolls. \citet{kvdirect2026} characterize this boundary on Gemma-3:
key reconstruction from the residual alone is inexact on the $29$ of $34$
windowed layers (values stay exact), and they identify the local position
index as the missing piece. We implement and verify that augmentation: extending the checkpoint from $h$ to $(h,p)$, which adds four
extra bytes per token ($<0.1\%$), and re-rotating with the layer's own base at index $p$
at reconstruction. On Gemma-3 (1B and 4B; on 4B, $29$ of $34$ layers windowed) this is bit-exact at
every windowed and global layer, at sequence lengths past the window
($T{=}1498$ vs.\ window $1024$; Table~\ref{tab:bitexact}), including with the
rolled cache layout, where every sliding layer retains only its last $W$ in-window
tokens and their $K/V$ are rebuilt from $(h,p)$ alone. This recovers the
residual-recompute option that \citet{slimattention2025} list as RoPE-incompatible.
Table~\ref{tab:complete} scopes the claim: the two RoPE cells are measured; the
remaining cells follow from where position enters the computation. NoPE and
learned absolute embeddings leave $K/V$ projections position-free or fold
position into $h$ itself, and ALiBi biases attention scores rather than $K$.
We have not measured these cases, and encoder--decoder cross-attention is out of
scope.

\begin{table}
\centering\papertablefont
\caption{State the checkpoint must carry for exact reconstruction.
``By construction'' cells are not measured.}
\label{tab:complete}
\setlength{\tabcolsep}{4.5pt}
\maxsizebox{\linewidth}{!}{%
\begin{tabular}{llll}
\toprule
attention & position encoding & checkpoint & status \\
\midrule
global & RoPE & $h$ & verified (Table~\ref{tab:bitexact}) \\
interleaved sliding-window & dual-base RoPE & $(h,p)$ & verified, Gemma-3 \\
any & NoPE \citep{nope2023} & $h$ & by construction \\
any & learned absolute & $h$ & by construction \\
any & ALiBi \citep{alibi2021} & $h$ & by construction \\
cross-attention & any & --- & out of scope \\
\bottomrule
\end{tabular}
}
\end{table}

Fused attention kernels reorder floating-point operations, so a bit-exactness
claim must be re-checked per backend. Table~\ref{tab:bitexact} reports
$\max|\Delta K|=\max|\Delta V|=0$ across CPU, CUDA, and Ascend NPU, six architecture families including GQA, the Qwen3 $k$-norm path,
and Gemma-3's interleaved sliding-window attention, at scales from 135M to 32B,
in both fp32 and bf16. The 8B model on the NPU stays exact across a cumulative
$630$ layer executions (each of its $36$ layers verified independently from the
anchor); the $64$-layer 32B model on the A100 reconstructs every
layer to the bit.
Bit-exactness in bf16 depends on one invariant. Because eviction is a contiguous prefix (\S\ref{sec:discuss}), replay reproduces
the original forward over that prefix with the \emph{same} chunk boundaries and
the same kernel, so every floating-point reduction is issued in the same order
and the projected $K/V$ are identical bit for bit. Our downstream control
confirms the dependence: re-chunking the same computation (full vs.\
prefix+continuation) changes the reduction order and shifts logits by up to
$1.25$, whereas reconstruction keeps the order and leaves them unchanged.

\begin{table}
\centering\papertablefont
\caption{Bit-exact reconstruction across backends, architectures, and
precisions. Gen.\ match: exact tokens over 30 steps.}
\label{tab:bitexact}
\maxsizebox{\linewidth}{!}{%
\begin{tabular}{lccccc}
\toprule
model & backend & dtype & $\max|\Delta K|$ & $\max|\Delta V|$ & gen. match \\
\midrule
SmolLM2-135M & CPU / CUDA & fp32 & 0 & 0 & 30/30 \\
Qwen2.5-0.5B (GQA) & CUDA & fp32/bf16 & 0 & 0 & 30/30 \\
Qwen3-0.6B ($k$-norm) & CUDA & bf16 & 0 & 0 & 30/30 \\
Qwen3-8B & Ascend 910B4 & bf16 & 0 & 0 & 30/30 \\
Qwen3-1.7B\,/\,4B & Ascend 910B4 & bf16 & 0 & 0 & 30/30 \\
Llama-3.2-1B\,/\,Llama-3.1-8B & Ascend 910B4 & bf16 & 0 & 0 & 30/30 \\
Qwen2.5-32B (GQA, $64$ layers) & A100 80GB & bf16 & 0 & 0 & 30/30 \\
Gemma-3-1B\,/\,4B (interleaved SWA) & Ascend 910B4 & bf16 & 0 & 0 & 30/30 \\
Mistral-7B-v0.3 & Ascend 910B4 & bf16 & 0 & 0 & 30/30 \\
\bottomrule
\end{tabular}
}
\end{table}

Table~\ref{tab:bitexact} covers architectures and backends. For
Mistral-7B-v0.3, its zero K/V maxima hold on all $32/32$ layers, and its
$30/30$ generation match means exact continuation logits and greedy tokens
under bf16 SDPA on Ascend 910B4.
The nine-model reconstruction protocol adds a balanced same-stack check: nine
reconstruction runs in fresh processes verify $262$ layerwise K/V reconstructions,
$270$ full-logit comparisons, and $270$ token continuations exactly.
Table~\ref{tab:reconstruction-validation} places the independently verified reconstruction
counts alongside the separate control and downstream protocols for each
checkpoint. Each protocol keeps its own column, so no steps, control arms, or
downstream traces are pooled.

\begin{table}
\centering\papertablefont
\caption{Reconstruction and downstream equivalence across nine models.
A: cascade/naive control (exact); B: direct residual reuse (expected nonzero
error).}
\label{tab:reconstruction-validation}
\maxsizebox{\linewidth}{!}{%
\begin{tabular}{@{}lrrrrrr@{}}
\toprule
 & \multicolumn{3}{c}{Reconstruction} & \multicolumn{2}{c}{Controls} & Downstream \\
\cmidrule(lr){2-4}\cmidrule(lr){5-6}\cmidrule(l){7-7}
model & exact K/V layers & exact logits & exact tokens & exact A & triggered B & exact traces \\
\midrule
Qwen2.5-0.5B & 24/24 & 30/30 & 30/30 & 4/4 & 4/4 & 49/49 \\
Qwen3-0.6B & 28/28 & 30/30 & 30/30 & 4/4 & 2/2 + 2 \textsc{N/A--Arch} & 49/49 \\
Qwen3-1.7B & 28/28 & 30/30 & 30/30 & 4/4 & 4/4 & 49/49 \\
Qwen3-4B & 36/36 & 30/30 & 30/30 & 4/4 & 4/4 & 49/49 \\
Qwen3-8B & 36/36 & 30/30 & 30/30 & 4/4 & 4/4 & 49/49 \\
Llama-3.2-1B-Instruct & 16/16 & 30/30 & 30/30 & 4/4 & 4/4 & 49/49 \\
Llama-3.1-8B-Instruct & 32/32 & 30/30 & 30/30 & 4/4 & 4/4 & 49/49 \\
SmolLM2-135M & 30/30 & 30/30 & 30/30 & 4/4 & 4/4 & 49/49 \\
Mistral-7B-v0.3 & 32/32 & 30/30 & 30/30 & 4/4 & 4/4 & 49/49 \\
\bottomrule
\end{tabular}
}
\end{table}

A paired control matrix then runs each of nine models at
$N\in\{256,4096\}$ and both $\sstar$ and $L$, for $18$ independent runs and $36$ arm rows.
Control A checks cascade replay against the exact naive-replay
reference; Control B requires direct residual reuse without replay to trigger a
nonzero-error negative control. Table~\ref{tab:reconstruction-validation} reports the outcomes separately for applicable and inapplicable settings. Control A is exact in all $36/36$ rows; Control
B triggers in all $34/34$ applicable rows, while Qwen3-0.6B's two $s^\star=1$
rows are \textsc{N/A--Arch} because every layer is anchored.

\subsection{Replay Efficiency and Anchor Placement}

\begin{figure}
\centering
\includegraphics[width=\linewidth]{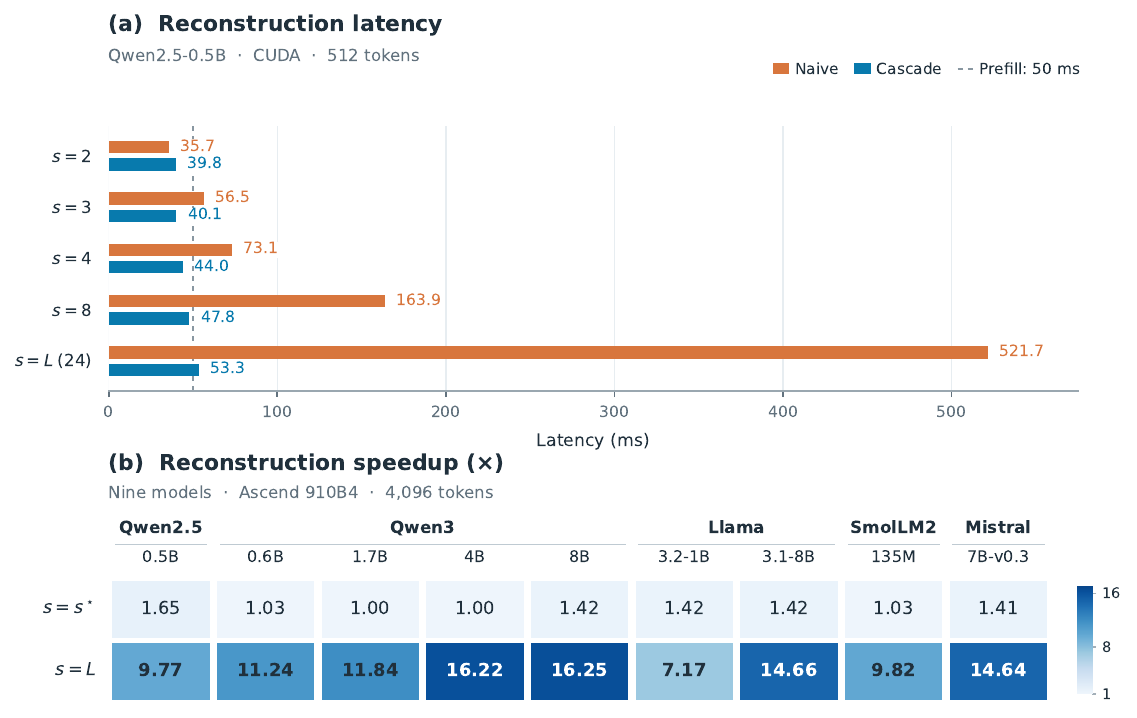}
\caption{\textbf{Cascade versus naive reconstruction.}
(a) Qwen2.5-0.5B on CUDA at $N{=}512$; dashed line: re-prefill.
(b) Speedup on Ascend 910B4 at $N{=}4096$.}
\Description{The upper panel compares naive and cascade reconstruction with
paired horizontal latency bars at five selected anchor spacings. A dashed
vertical line marks re-prefill latency. The lower panel is a two-by-nine
speedup matrix, with models grouped by family; its two rows show the
break-even and deepest spacings. Each colored cell includes its numeric value.}
\label{fig:gate}
\end{figure}

Figure~\ref{fig:gate} compares cascade with independent per-layer replay.
Naive per-layer replay climbs to $522$\,ms at $s=L$ while
cascade replay stays near-flat around $40$--$53$\,ms, $9.8\times$ faster, and
both remain bit-exact. We report speedup as naive-replay latency divided by
cascade-replay latency. The matched nine-model $4$k controls reproduce the
same depth-amplified gap: $s=L$ yields $7.17\times$--$16.25\times$ speedup
(median $11.84\times$), whereas $s=\sstar$ stays near unity (median
$1.41\times$). Cascade and naive replay are exact in all $36/36$ rows. All
latencies here and below are medians of $5$ device-synchronized repetitions
after warmup.

\begin{figure}
\centering
\includegraphics[width=\linewidth]{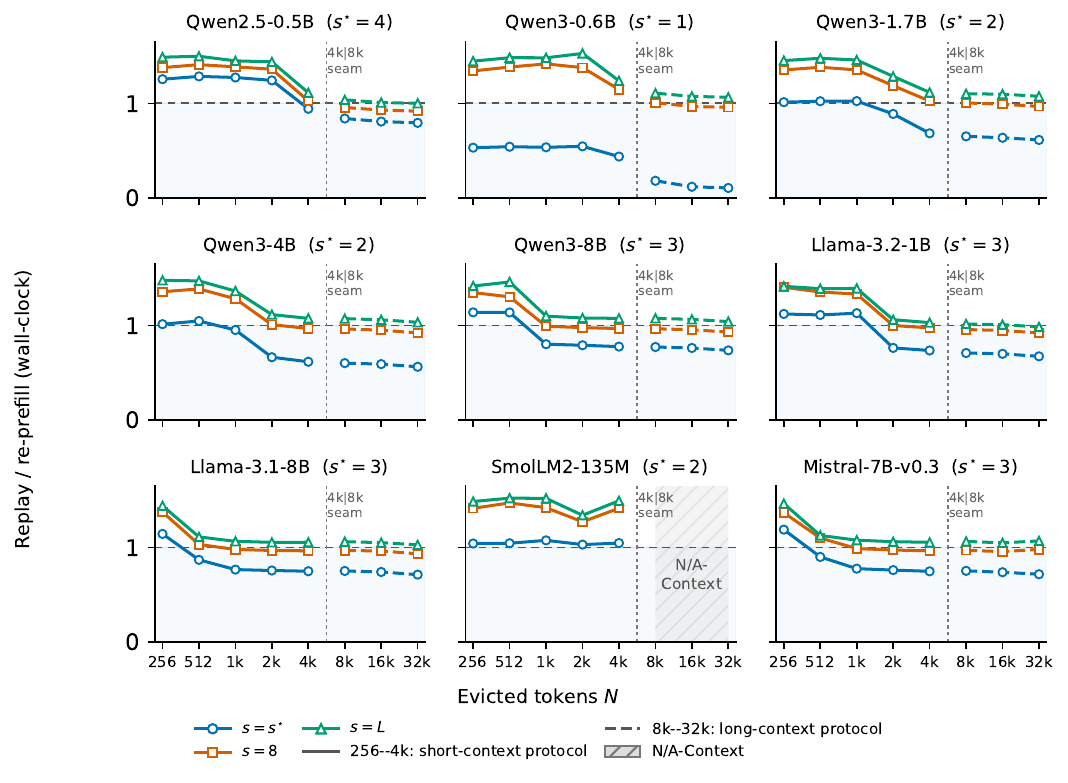}
\caption{\textbf{Replay/re-prefill latency across nine models.} Solid
($256$--$4$k) and dashed ($8$k--$32$k) segments come from different protocols.
At nominal $32$k, Qwen2.5 and Mistral use $32764$ prompt tokens.}
\Description{Nine small-multiple panels, one per checkpoint, plot
replay divided by re-prefill time at three anchor spacings. Solid short-context
curves end at 4k, dashed long-context curves begin at 8k, and the protocol seam
is marked without connecting the two series. The panels contain $207$ valid
endpoints and explicitly mark nine SmolLM2 \textsc{N/A--Context} endpoint slots.}
\label{fig:crossmodel-context}
\end{figure}

\begin{table}
\centering\papertablefont
\caption{Peak allocated memory (GiB). Short- and long-context columns come
from different protocols; context lengths as in Fig.~\ref{fig:crossmodel-context}.}
\label{tab:crossmodel-peak-memory}
\begin{tabular*}{\linewidth}{@{}l@{\extracolsep{\fill}}rrrrrrrr@{}}
\toprule
 & \multicolumn{5}{c}{short context} & \multicolumn{3}{c}{long context} \\
\cmidrule(lr){2-6}\cmidrule(l){7-9}
model / peak GiB & 256 & 512 & 1k & 2k & 4k & 8k & 16k & 32k \\
\midrule
Qwen2.5-0.5B & 0.97 & 1.00 & 1.06 & 1.12 & 1.31 & 1.36 & 1.80 & 2.69 \\
Qwen3-0.6B & 1.25 & 1.38 & 1.63 & 2.14 & 3.14 & 2.64 & 4.11 & 7.06 \\
Qwen3-1.7B & 3.36 & 3.50 & 3.78 & 4.34 & 5.45 & 4.90 & 6.59 & 9.97 \\
Qwen3-4B & 7.70 & 7.89 & 8.27 & 9.03 & 10.54 & 9.89 & 12.28 & 17.07 \\
Qwen3-8B & 15.49 & 15.71 & 16.14 & 17.00 & 18.72 & 17.89 & 20.51 & 25.77 \\
Llama-3.2-1B-Instruct & 2.37 & 2.42 & 2.52 & 2.71 & 3.10 & 3.18 & 4.06 & 5.81 \\
Llama-3.1-8B-Instruct & 15.17 & 15.36 & 15.75 & 16.52 & 18.05 & 17.49 & 20.03 & 25.10 \\
SmolLM2-135M & 0.30 & 0.33 & 0.40 & 0.52 & 0.76 & \textsc{N/A--C} & \textsc{N/A--C} & \textsc{N/A--C} \\
Mistral-7B-v0.3 & 13.71 & 13.90 & 14.29 & 15.06 & 16.59 & 16.04 & 18.57 & 23.64 \\
\bottomrule
\end{tabular*}

\end{table}

\begin{figure}
\centering
\includegraphics[width=0.82\linewidth]{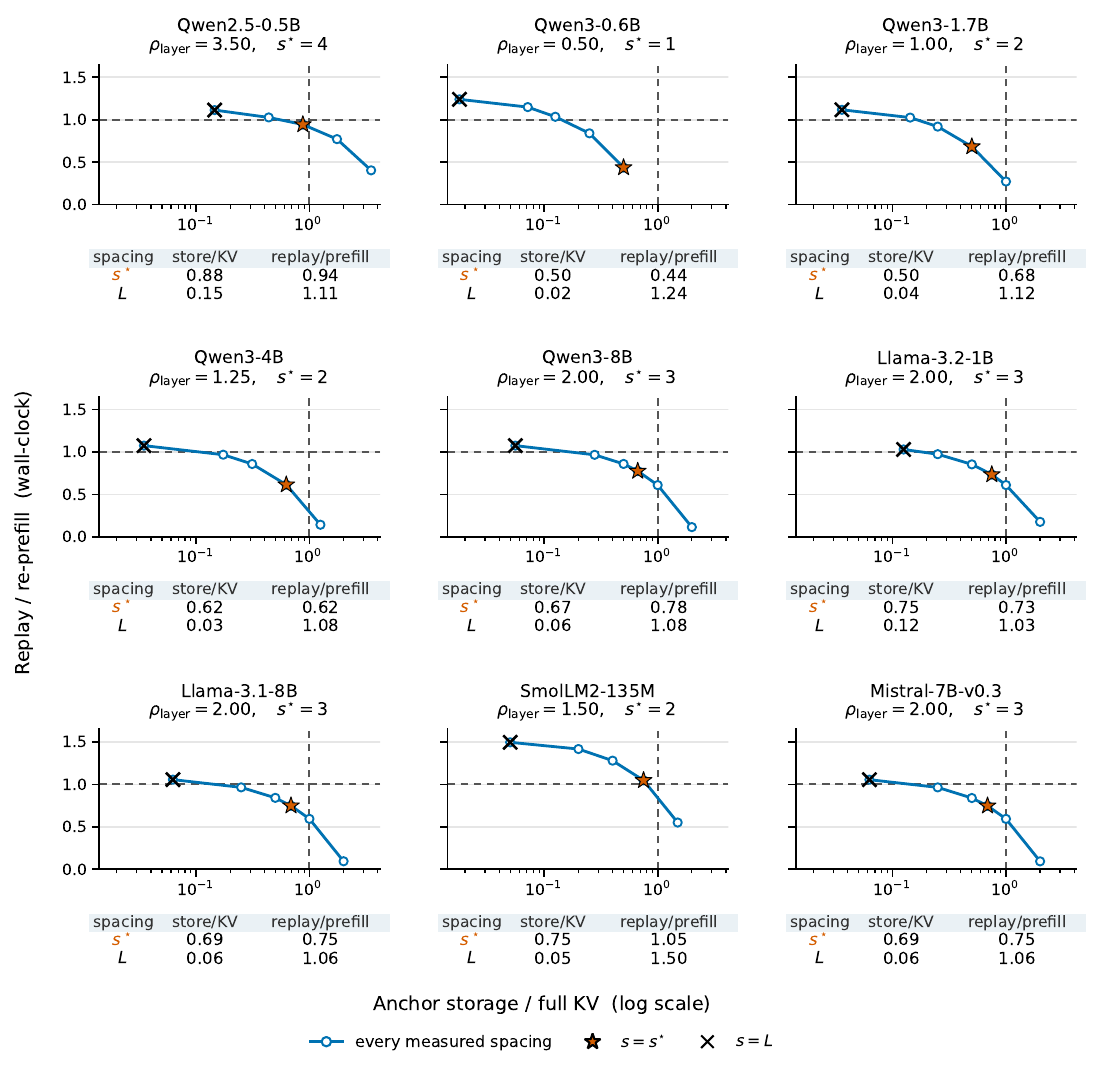}
\caption{\textbf{Measured $4$k storage--replay frontier across nine models on Ascend.}
Dashed lines mark full-KV storage and one re-prefill; stars mark $\sstar$,
crosses $s=L$. Timing covers reconstruction only.}
\Description{Nine small-multiple panels show all measured four-thousand-token
storage and replay operating points. Dashed reference lines mark equal storage
to full KV and equal time to re-prefill; stars highlight each model's
break-even spacing and crosses its deepest endpoint. Each panel includes the
architecture ratio and a two-row numerical summary of those endpoints.}
\label{fig:measured-frontier}
\end{figure}

The measured points in Figure~\ref{fig:measured-frontier} define deployment choices. A memory-constrained deployment may use the first
storage-saving point $\sstar$; a latency-sensitive deployment may choose a
shallower spacing, while a capacity-oriented deployment may accept the replay
cost of a deeper spacing. For non-uniform demand, the scalar spacing is replaced
by the anchor set selected by the exact placement procedure.

\subsection{Capacity and Session Switching}\label{sec:capacity}

We now test the predicted memory savings against measured out-of-memory (OOM)
boundaries. The setting is multi-session serving: many long-context sessions
resident on the accelerator, one active at a time. Inactive sessions are evicted
down to their anchors; reactivating one cascades the replay above to rebuild its
full cache. We prefill real $32$k-token sessions one at a time until the device
throws out-of-memory, recording two numbers per configuration:
\emph{resident}, the session count at the OOM boundary, and \emph{usable}, the
largest count at which the last session can still be reactivated in place, with
reconstructed KV bit-identical to a retained ground truth and four greedy
continuation steps reproducing the full-cache logits exactly. Every
reactivation that fit in memory met both conditions
($\Delta K=\Delta\text{logits}=0$).

\begin{figure}
\centering
\includegraphics[width=\linewidth]{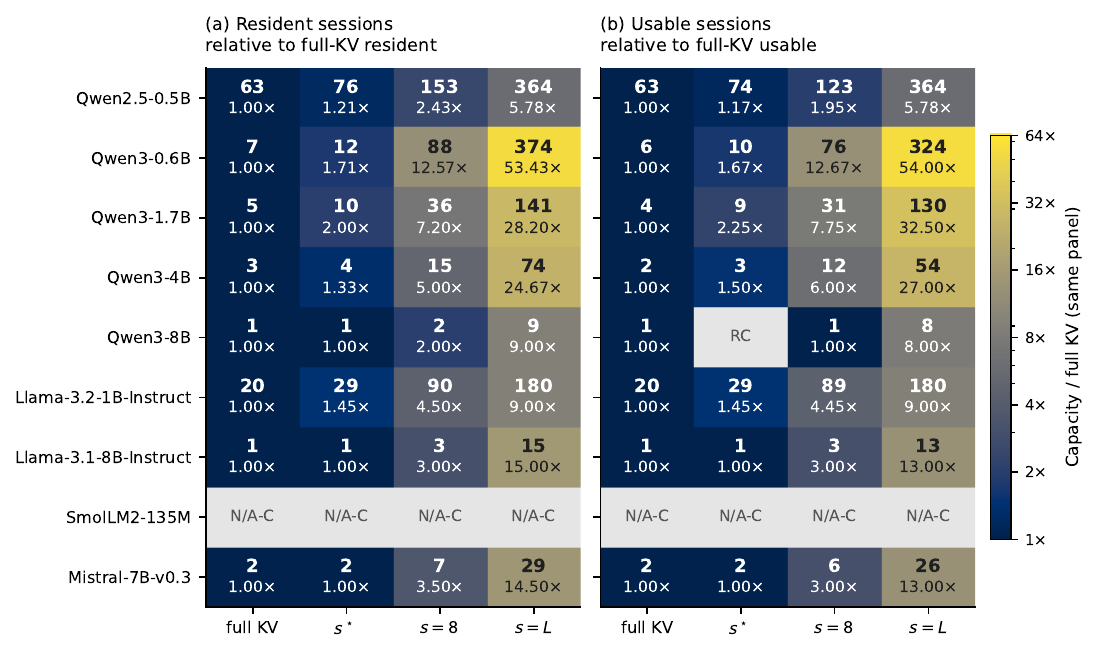}
\caption{\textbf{Session capacity at $32$k on Ascend.} Cells give session
counts and multipliers over the matching full-KV capacity (resident left,
usable right). Qwen3-8B at $s=\sstar$ is resource-censored.}
\Description{Two nine-row heatmaps show resident and operationally usable
capacity as absolute session counts and multipliers relative to the matching
full-KV boundary for four policies. All nine checkpoints
are displayed; context-inapplicable and resource-censored cells are marked.}
\label{fig:capacity}
\end{figure}

The primary capacity evidence is the controlled nine-model panel on one
Ascend software and hardware stack. Eight global-attention checkpoints
support a $32$k context with the required continuation and contribute
$8\times4=32$ measured policy points; SmolLM2 is explicitly
\textsc{N/A--Context}. Figure~\ref{fig:capacity} reports the
absolute resident and usable counts in each cell together with the multiplier
relative to the corresponding full-KV boundary.
At $s=\sstar$, usable capacity ranges from resource-censored to $2.25\times$
full KV, and two models tie the baseline. At $s=8$, the range is
$1.00\times$--$12.67\times$; at $s=L$, every applicable checkpoint improves,
by $5.78\times$--$54.0\times$. Resource-censored points remain marked as such.

Operational serving break-even also depends on active-slot materialization during
prefill or reactivation. The capacity multiplier appears consistently at deep
anchors, which cascade replay makes affordable. The $s=L$ endpoint stores one
$d$-vector per evicted token, matching the single-vector residual budget of
KV-Direct (Table~\ref{tab:endpoints}). Beyond that endpoint, the spectrum adds
the measured interior, which trades some of that capacity for lower replay work.

\subsubsection{Serving under Session Switching}\label{sec:trace}

To test whether resident sessions can be served, we run a session-switching
trace: $N$ long-context sessions served
round-robin, each activation continuing the same prefix by $32$ greedy tokens
before yielding. Every continuation is compared token-by-token, and its
final-step logits are compared to a never-evicted ground truth. Two
policies face the identical load: full KV (LRU evict and re-prefill beyond its
measured usable cap) and anchors (device-only). TTFT is reactivation plus the first
decode step; all decoding is greedy.

For Qwen2.5, the serving prefix is $C=32{,}736$ tokens, and the session
load uses a capacity calibration at the same length. Its $C=32{,}768$ capacity
measurement in Fig.~\ref{fig:capacity} is reported separately.

\begin{table}
\centering\papertablefont
\caption{Session-switching performance: two rounds of 32 generated tokens per
visit. Throughput includes reactivation and reconstruction, so it is not
steady-state decode throughput. Qwen3-8B at $s=\sstar$ is resource-censored.}
\label{tab:crossmodel-serving}
\setlength{\tabcolsep}{3.5pt}
\maxsizebox{\linewidth}{!}{%
\begin{tabular}{@{}llrccr@{}}
\toprule
model & policy & $N$ & TTFT p50 (ms) & TTFT p99 (ms) & 32-token service tok/s \\
\midrule
Qwen2.5-0.5B & full KV & 66 & 48.9 & 73.1 & 23.6489 \\
 & $s=s^\star=4$ & 80 & 990.9 & 997.0 & 13.7504 \\
 & $s=8$ & 161 & 1137.6 & 1147.1 & 12.9320 \\
 & $s=L$ & 484 & 1239.7 & 1250.3 & 12.2923 \\
\midrule
Qwen3-0.6B & full KV & 6 & 104.5 & 144.6 & 16.0851 \\
 & $s=s^\star=1$ & 10 & 297.1 & 350.6 & 14.1546 \\
 & $s=8$ & 76 & 1859.2 & 1881.1 & 8.5114 \\
 & $s=L$ & 324 & 2053.3 & 2076.7 & 3.4381 \\
\midrule
Qwen3-1.7B & full KV & 4 & 93.3 & 168.7 & 15.7532 \\
 & $s=s^\star=2$ & 9 & 1482.8 & 1529.8 & 9.2494 \\
 & $s=8$ & 31 & 2282.5 & 2301.4 & 7.5832 \\
 & $s=L$ & 130 & 2525.3 & 2544.2 & 7.1683 \\
\midrule
Qwen3-4B & full KV & 2 & 135.5 & 248.9 & 8.0434 \\
 & $s=s^\star=2$ & 3 & 3216.3 & 3254.5 & 4.5502 \\
 & $s=8$ & 12 & 5185.2 & 5215.6 & 3.5570 \\
 & $s=L$ & 54 & 5788.5 & 5808.2 & 3.3373 \\
\midrule
Qwen3-8B & full KV & 1 & 150 & -- & 7.46 \\
 & $s=s^\star=3$ & \textsc{RC} & \textsc{RC} & -- & \textsc{RC} \\
 & $s=8$ & 1 & 6310 & -- & 3.06 \\
 & $s=L$ & 8 & 7060 & -- & 2.86 \\
\midrule
Llama-3.2-1B-Instruct & full KV & 20 & 39.3 & 47.8 & 34.1448 \\
 & $s=s^\star=3$ & 29 & 1223.6 & 1228.9 & 15.1582 \\
 & $s=8$ & 89 & 1661.3 & 1678.1 & 12.5515 \\
 & $s=L$ & 180 & 1773.0 & 1785.5 & 11.9587 \\
\midrule
Llama-3.1-8B-Instruct & full KV & 1 & 130.6 & 187.4 & 8.2901 \\
 & $s=s^\star=3$ & 1 & 4439.2 & 4473.1 & 3.9304 \\
 & $s=8$ & 3 & 5774.1 & 5783.6 & 3.3788 \\
 & $s=L$ & 13 & 6332.7 & 6374.5 & 3.1905 \\
\midrule
SmolLM2-135M & -- & \textsc{N/A-C} & \textsc{N/A-C} & \textsc{N/A-C} & \textsc{N/A-C} \\
\midrule
Mistral-7B-v0.3 & full KV & 2 & 130.1 & 176.2 & 8.4188 \\
 & $s=s^\star=3$ & 2 & 4865.2 & 4865.9 & 3.7842 \\
 & $s=8$ & 6 & 6214.4 & 6273.4 & 3.2516 \\
 & $s=L$ & 26 & 6763.6 & 6787.9 & 2.9304 \\
\bottomrule
\end{tabular}

}
\end{table}

Table~\ref{tab:crossmodel-serving} is the complete modelwise serving evidence:
all eight context-applicable checkpoints are visible under full KV,
$s=\sstar$, $s=8$, and $s=L$, while SmolLM2 remains explicitly
\textsc{N/A--Context}. In separate capacity-pressure tests, Qwen3-1.7B,
Qwen3-4B, and Llama-3.2-1B complete their full-KV $F{+}1$ points at
$N{=}5,3,21$, respectively, by miss-triggered re-prefill; all $58$ scheduled
visits ($1{,}856$ generated tokens) and final logits are exact. Llama-3.1-8B
at $F{+}1{=}2$ instead completes one exact visit ($32/32$ tokens and final
logits) before the trace is truncated, so that point supports only a bit-exact
partial trace. It is the only truncation among the capacity-pressure runs,
and all $45{,}600$ observed tokens are exact.
The nine-model matrix remains the primary serving evidence.

To test whether the timing conclusions depend on one endpoint observation, we
repeat the $s{=}\sstar$ and $s{=}L$ spacing endpoints at fixed session loads across the applicable global-attention models.
The experiment contains $45$ performance measurements in independent processes
across $15$ eligible model--endpoint groups, plus one
Qwen3-8B resource-censored group. Table~\ref{tab:endpoint-repeats}
reports complete groups; the censored group is displayed as \textsc{RC}.
The table distinguishes measured full-KV capacity $F$, usable anchor capacity
$U$, and the actual repetition load $N$. Some repetitions use $N<U$:
for Qwen3-1.7B at $s=L$, for example, $N=5$ although $U=130$ and the two-round
serving row uses $N=130$. These repetitions assess stability at their stated
loads; the two-round serving rows use their own loads.

\begin{table}
\centering\papertablefont
\caption{Repeated endpoint performance, three processes per group. $F$, $U$:
full-KV and usable anchor capacity; $N$: tested load.}
\label{tab:endpoint-repeats}
\maxsizebox{\linewidth}{!}{%
\begin{tabular}{@{}llrrrrrr@{}}
\toprule
model & policy & $F$ & $U$ & $N$ & reps & TTFT p50 ms & tok/s \\
\midrule
Qwen2.5-0.5B & $s=s^\star$ & 63 & 74 & 64 & 3/3 & 949.5 & 14.14 \\
 & $s=L$ & 63 & 364 & 64 & 3/3 & 1186.4 & 12.63 \\
\midrule
Qwen3-0.6B & $s=s^\star=1$ & 6 & 10 & 10 & 3/3 & 297.1 & 14.269 \\
 & $s=L$ & 6 & 324 & 324 & 3/3 & 2053.3 & 3.462 \\
\midrule
Qwen3-1.7B & $s=s^\star$ & 4 & 9 & 5 & 3/3 & 1463.5 & 9.45 \\
 & $s=L$ & 4 & 130 & 5 & 3/3 & 2511.3 & 7.20 \\
\midrule
Qwen3-4B & $s=s^\star$ & 2 & 3 & 3 & 3/3 & 3213.3 & 4.54 \\
 & $s=L$ & 2 & 54 & 3 & 3/3 & 5789.4 & 3.33 \\
\midrule
Qwen3-8B & $s=s^\star$ & 1 & 0 & \textsc{RC} & 0/3 & \textsc{RC} & \textsc{RC} \\
 & $s=L$ & 1 & 8 & 2 & 3/3 & 7058.4 & 2.86 \\
\midrule
Llama-3.2-1B-Instruct & $s=s^\star$ & 20 & 29 & 21 & 3/3 & 1218.5 & 15.19 \\
 & $s=L$ & 20 & 180 & 21 & 3/3 & 1768.4 & 12.00 \\
\midrule
Llama-3.1-8B-Instruct & $s=s^\star$ & 1 & 1 & 1 & 3/3 & 4448.5 & 3.92 \\
 & $s=L$ & 1 & 13 & 2 & 3/3 & 6359.8 & 3.18 \\
\midrule
SmolLM2-135M & $s=s^\star$ & \textsc{N/A-C} & \textsc{N/A-C} & \textsc{N/A-C} & 0/0 & \textsc{N/A-C} & \textsc{N/A-C} \\
 & $s=L$ & \textsc{N/A-C} & \textsc{N/A-C} & \textsc{N/A-C} & 0/0 & \textsc{N/A-C} & \textsc{N/A-C} \\
\midrule
Mistral-7B-v0.3 & $s=s^\star=3$ & 2 & 2 & 2 & 3/3 & 4810.0 & 3.803 \\
 & $s=L$ & 2 & 26 & 26 & 3/3 & 6807.0 & 2.927 \\
\bottomrule
\end{tabular}
}
\end{table}

For Qwen2.5-0.5B, Qwen3-1.7B, Qwen3-4B, Qwen3-8B,
Llama-3.2-1B-Instruct, and Llama-3.1-8B-Instruct, the largest three-repeat
range across the eligible endpoint groups is
$0.59\%$ of the median for TTFT and $4.24\%$ for throughput. The four endpoint groups for Qwen3-0.6B and Mistral have mean TTFT/throughput of $0.297$\,s/$14.269$ token/s and
$2.053$\,s/$3.462$ token/s for Qwen3-0.6B at $s^\star/L$, and
$4.810$\,s/$3.803$ token/s and $6.807$\,s/$2.927$ token/s for Mistral.
Qwen3-8B's $s^\star$ group remains resource-censored and contributes no timing
aggregate.

\subsubsection{Evaluation across Nine Global-Attention Models}
\label{sec:nine-model-evaluation}

We evaluate Qwen2.5-0.5B; Qwen3-0.6B, Qwen3-1.7B, Qwen3-4B, and
Qwen3-8B; Llama-3.2-1B-Instruct and Llama-3.1-8B-Instruct; SmolLM2-135M;
and Mistral-7B-v0.3 under the same global-attention evaluation protocols. Nine independent
reconstruction runs and $45$ model--context timing runs verify all $262$
layer checks, $270$ full-logit comparisons, $270$ token continuations, and
$245$ timing rows exactly. Separate control and downstream protocols add
$18$ independent control runs ($36$ cascade rows) and $9\times49=441$ exact downstream
traces. The long-context protocol contributes $24$ independent runs and $72$ timing rows.
SmolLM2 cannot satisfy the $8192{+}4$ token requirement and is therefore
\textsc{N/A--Context}; Qwen3-0.6B's direct-reuse negative control at
$s^\star=1$ is \textsc{N/A--Arch} because there is no non-anchor layer.

The eight context-applicable checkpoints each contribute four capacity and
four two-round session-switching measurements. Seven checkpoints complete all $28$ serving
measurements without OOM; Qwen3-8B contributes four additional
measurements, including the resource-censored $s^\star$ point. Endpoint
repeatability is measured in three fresh processes for each eligible
model--endpoint group, yielding $45$ successful repetitions across $15$ groups
and one explicitly resource-censored group. Performance summaries use complete attempts. Figure~\ref{fig:capacity} and
Tables~\ref{tab:crossmodel-serving}, \ref{tab:endpoint-repeats},
and~\ref{tab:reconstruction-validation} report these distinct measurements
without pooling protocols or counting the same setting twice.

\subsubsection{Native Gemma-3 Serviceable Serving}\label{sec:gemma-serviceable}

The interleaved Gemma-3 cache requires a separate operational calibration. Its
inactive state retains every sliding-window layer's live window, the absolute
position vector and cursor, and residual anchors only for reconstructing the
full-history global layers. Consequently, the dense-equivalent $\sstar$ in
Fig.~\ref{fig:spectrum} is not the native resident break-even. At the requested
$32$k point, the native resident rule gives $s^\star_{\rm nat}{=}26$ for
Gemma-3-1B and $12$ for Gemma-3-4B. For 1B, this is $s^\star_{\rm nat}=L=26$
because 22 local-window layers remain resident and only four global layers are
replaced by residual anchors. The threshold uses the same byte comparison as
the dense audit, but only the global full-history KV is replaced. The native
sliding-window KV and position state are retained by both alternatives and
therefore cancel in the strict comparison. For Gemma-3-1B,
$G=4$, $d=1152$, $n_{kv}=1$, $d_{\rm head}=256$, and bf16 gives
$4\times(2\cdot1\cdot256\cdot2)=4096$ bytes per token of global KV, while
one residual anchor costs $1152\cdot2=2304$ bytes. At $s=13$, two anchors
cost $4608$ bytes and only tie or exceed the global KV budget; at $s=26$, one
anchor costs $2304$ bytes, so $s^\star_{\rm nat}=26=L$. For Gemma-3-4B,
$G=5$, $d=2560$, $n_{kv}=4$, and $d_{\rm head}=256$, so global KV costs
$5\times(2\cdot4\cdot256\cdot2)=20480$ bytes per token and one residual
anchor costs $2560\cdot2=5120$ bytes. Four anchors at $s=11$ tie this budget,
whereas three anchors at $s=12$ cost $15360$ bytes; hence
$s^\star_{\rm nat}=12$. These native thresholds are uniform-spacing storage
thresholds and are distinct from the DP budget, which places anchors on the
global demand layers. The native placement
check above uses $k=4$ for 1B and $k=5$ for 4B, placing anchors at the global
demand layers and achieving zero replay with bit-exact K/V and continuation
outputs. Tables~\ref{tab:gemma-longcontext} and
\ref{tab:gemma-serviceable-summary}(a,b) retain uniform-spacing policies for
their resident and capacity-bound measurements. Panel (c) adds a separate
fixed-budget explicit-anchor service trace: at the same $k$, uniform and DP
retain the same cold-state bytes, while DP reduces reactivation work. The trace
uses two cold states, two rounds, and four visits per arm. The same panel also
reports representative reactivation times against full KV and two targeted
explicit-anchor capacity boundaries.

The native long-context and residency evaluation comprises six long-context and eight resident-capacity
experiments. All $14$ measurements pass the numerical and cache-state
validation checks. The long-context measurements use four continuation
steps; multi-session serviceability is evaluated separately below with
$32$ steps per visit.

For all 18 model--context--spacing settings in
Table~\ref{tab:gemma-longcontext}, global-layer K/V, prompt logits, and each of
the four continuation logits and tokens are bit-exact.

\begin{table}
\centering\papertablefont
\caption{Native Gemma-3 long-context timing on Ascend 910B4. Replay cells give
median ms and replay/prefill ratio; memory is peak allocated GiB. For 1B,
$s^\star_{\rm nat}=L=26$.}
\label{tab:gemma-longcontext}
\setlength{\tabcolsep}{4.5pt}
\maxsizebox{\linewidth}{!}{%
\begin{tabular}{@{}lrrrrrr@{}}
\toprule
model & requested / actual $N$ & prefill (ms) & $s^\star_{\rm nat}$ replay & $s{=}8$ replay & $s{=}L$ replay & peak (GiB) \\
\midrule
\multirow{3}{*}{Gemma-3-1B}
 & $8192/8192$   & 3093.298  & 2915.465 / 0.9425 & 2187.738 / 0.7073 & -- & 2.258 \\
 & $16384/16384$ & 6203.260  & 5893.023 / 0.9500 & 4474.127 / 0.7213 & -- & 2.506 \\
 & $32768/32764$ & 12501.857 & 11861.298 / 0.9488 & 9026.867 / 0.7220 & -- & 2.999 \\
\midrule
\multirow{3}{*}{Gemma-3-4B}
 & $8192/8192$   & 4127.417  & 3806.159 / 0.9222 & 2976.811 / 0.7212 & 3710.724 / 0.8990 & 9.291 \\
 & $16384/16384$ & 8651.576  & 7476.372 / 0.8642 & 6083.218 / 0.7031 & 7710.674 / 0.8912 & 10.220 \\
 & $32768/32768$ & 18712.117 & 16260.324 / 0.8690 & 12909.363 / 0.6899 & 16978.216 / 0.9073 & 12.074 \\
\bottomrule
\end{tabular}

}
\end{table}

Table~\ref{tab:gemma-serviceable-summary}, panel (a), joins each measured resident bound to
the corresponding calibrated serviceable working point. $R_{\rm res}$ is the last
resident count before an observed $R_{\rm res}+1$ OOM, verified by one active-slot
restore and four exact decode steps. Serviceability calibration starts from this fixed count and the
$\lfloor0.90R_{\rm res}\rfloor$ candidate. A candidate becomes $U_{\rm svc}$
only after a fresh process builds all inactive
states, restores one native interleaved cache, and completes $32$ greedy decode
steps with every token and full-logit comparison exact. All eight calibration
runs passed; their timing sets the operating points and is omitted from performance means.
Panel (a) reports the eight two-round serving traces at those fixed
operating points; panel (b) summarizes fresh-process repetition groups at each
spacing endpoint. The equivalent 1B $s^\star_{\rm nat}$ and $s=L$ groups are
pooled over six processes, while the distinct 4B endpoints retain three
processes each. Serving-trace p50 values and the mean/SD of repeated p50 values
remain distinct statistics.

Serving uses 32,736-token prefixes for 1B and 32,768-token prefixes for
4B. At load $U_{\rm svc}$, two rounds contain $2U_{\rm svc}$ visits and
$64U_{\rm svc}$ generated tokens; every token and corresponding full-logit
comparison is exact. Each endpoint repetition has the same allocated and
reserved memory peaks as its corresponding serving row.

Because Gemma-3-1B has $s^\star_{\rm nat}=s=L=26$, the table reports one
canonical merged endpoint row, with the duplicate $s=L$ repetitions included
in the pooled endpoint statistics. Gemma-3-4B retains separate $s^\star_{\rm nat}=12$
and $s=L=34$ rows because they are different policies.

\begin{table}
\centering\papertablefont
\caption{Native Gemma-3 capacity and serving on Ascend 910B4 at nominal 32k
context. (a) Resident bound $R_{\rm res}$ and serviceable load $U_{\rm svc}$;
A/R: allocated/reserved memory. (b) TTFT p50 and service tok/s, mean $\pm$ SD
over repeated processes (six for 1B, three for 4B). (c) Explicit-anchor
uniform vs.\ DP at a 32,736-token prefix.}
\label{tab:gemma-serviceable-summary}
\begingroup
\setbox0=\hbox{%
\maxsizebox{\linewidth}{!}{%
\begin{tabular}{@{}llrrrrrrr@{}}
\toprule
\multicolumn{9}{@{}l}{\textit{(a) Resident bounds and calibrated serviceable operating points}} \\
model & arm & $s$ & \shortstack{state\\MiB} & $R_{\rm res}\!\rightarrow\!U_{\rm svc}$ & \shortstack{TTFT p50\\(ms)} & \shortstack{32-token service\\tok/s} & \shortstack{Resident peak A/R\\(GiB)} & \shortstack{Serving peak A/R\\(GiB)} \\
\midrule
Gemma-3-1B & $s^\star_{\rm nat}=s{=}L$ & 26 & 83.179 & $332\!\rightarrow\!298$ & 12541.500 & 2.0604 & 28.974 / 29.008 & 26.671 / 26.729 \\
 & $s{=}8$ & 8 & 298.969 & $92\!\rightarrow\!82$ & 8835.433 & 2.7280 & 28.861 / 28.924 & 26.401 / 26.469 \\
\midrule
Gemma-3-4B & $s^\star_{\rm nat}$ & 12 & 596.250 & $34\!\rightarrow\!30$ & 16124.455 & 1.6052 & 28.547 / 28.605 & 27.618 / 27.828 \\
 & $s{=}8$ & 8 & 916.250 & $22\!\rightarrow\!19$ & 12585.676 & 1.9624 & 28.435 / 28.484 & 27.151 / 27.359 \\
 & $s{=}L$ & 34 & 276.250 & $75\!\rightarrow\!67$ & 17004.657 & 1.5377 & 28.984 / 29.064 & 28.351 / 28.463 \\
\bottomrule
\end{tabular}%
}%
}%
\usebox0
\par\smallskip
\begin{tabular*}{\wd0}{@{}l@{\extracolsep{\fill}}lrrrr@{}}
\toprule
\multicolumn{6}{@{}l}{\textit{(b) Fresh-process endpoint repetitions (mean $\pm$ sample SD)}} \\
model & endpoint & $U_{\rm svc}$ & TTFT p50 (ms) & tok/s & CV TTFT / tok/s \\
\midrule
Gemma-3-1B & $s^\star_{\rm nat}=s{=}L$ & 298 & $11924.739 \pm 393.206$ & $2.147055 \pm 0.062031$ & 3.297\% / 2.889\% \\
\midrule
Gemma-3-4B & $s^\star_{\rm nat}$ & 30 & $15943.903 \pm 275.429$ & $1.620636 \pm 0.028944$ & 1.727\% / 1.786\% \\
 & $s{=}L$ & 67 & $17224.316 \pm 199.584$ & $1.512701 \pm 0.014899$ & 1.159\% / 0.985\% \\
\bottomrule
\end{tabular*}%
\par\smallskip
\begin{tabular*}{\wd0}{@{}l@{\extracolsep{\fill}}lrrrr@{}}
\toprule
\multicolumn{6}{@{}l}{\textit{(c) Representative Uniform--DP serviceability and targeted capacity}} \\
model & budget & policy & \shortstack{reactivation\\(s)} & \shortstack{Uniform / DP\\speedup} & \shortstack{capacity boundary\\last resident $\rightarrow$ first OOM} \\
\midrule
Gemma-3-1B & $k=1$ & DP (ours) & 14.63 & 1.14x & 322$\rightarrow$323 \\
 & $k=1$ & Uniform & 16.64 & -- & 322$\rightarrow$323 \\
\midrule
Gemma-3-4B & $k=3$ & DP (ours) & 17.38 & 1.16x & 31$\rightarrow$32 \\
 & $k=3$ & Uniform & 20.09 & -- & 31$\rightarrow$32 \\
\bottomrule
\end{tabular*}%
\papertablefont
\par
\endgroup

\end{table}

The native result links storage capacity to reactivation work in \S\ref{sec:capacity}.
Relative to full KV, $U_{\rm svc}$ is $1.67\times$ at the 1B native break-even
and $1.25\times$ at the 4B native break-even, while $s{=}8$ is still below full
KV ($0.46\times$ and $0.79\times$). The 4B deepest endpoint reaches
$67/24=2.79\times$. This capacity comes with replay cost: anchor TTFT is
$8.84$--$12.54$\,s for 1B versus $0.102$\,s for full KV, and
$12.59$--$17.00$\,s for 4B versus $0.137$\,s. Native anchoring therefore
extends serviceable capacity at a reactivation-latency cost. These ratios
measure cache reactivation only; end-to-end serving acceleration would require
a separate system measurement.

The targeted explicit-anchor capacity checks in panel (c) give the same
boundary for uniform and DP at the same budget: $322\to323$ OOM for 1B
$k=1$ and $31\to32$ OOM for 4B $k=3$. These points use the explicit-anchor
capacity protocol and validate equal resident bytes.

Across calibration, serving, and repetitions, all $28/28$ independent runs pass
(eight calibration, eight serving, and twelve repeated measurements),
without OOM. The three stages contribute $256$, $63{,}808$, and
$133{,}056$ exact token comparisons, respectively, and the same number of
full-logit comparisons; the maximum difference is zero. These measurements cover
the two native Gemma-3 checkpoints. The explicit-anchor DP trace in panel (c)
adds $20$ arms and $80$ exact visits ($2{,}560$ token and full-logit comparisons);
it is a fixed two-session service trace reported separately from the capacity and
endpoint summaries. Preliminary resource probes set protocol parameters and are
omitted from performance averages.

\subsection{Downstream Equivalence}

Under the matched execution conditions of \S\ref{sec:replay}, bit-exact KV
implies bit-exact continuation logits, which implies every
logit-derived downstream metric is unchanged. We verify reconstruction and continuation on
benchmark data: prefill with hidden states, evict all but the anchors, cascade
reconstruct, then continue decoding from both the true and reconstructed caches.
In the representative Qwen2.5-0.5B downstream evaluation, the reconstructed-cache
continuation logits are bitwise identical to the true cache over $1765$
examples spanning six tasks plus long-context generation
($\Delta=0$ throughout), and $121$ greedy generation sequences of $32$ steps each
show zero token mismatch. As a control, changing only the forward chunking of
the \emph{same} computation (full vs. prefix+continuation, a bf16 kernel effect)
shifts logits by up to $1.25$; against that noise floor, replacing the true cache
with the reconstructed one shifts them by exactly zero.

The nine-model comparison evaluates $49$ traces per checkpoint, using
samples from seven tasks. All $441$ traces reproduce K/V, logits, log-likelihoods, and
greedy tokens exactly. Table~\ref{tab:reconstruction-validation} reports the results for each model separately from the more detailed
single-model evaluation.

The evaluation includes $72$ traces for each of HellaSwag, PIQA,
ARC-Easy, ARC-Challenge, WinoGrande, and LAMBADA, plus $9$ long-generation
traces from WikiText. The $576$ generated continuation tokens and every scored
log-likelihood remain exact.

Because the outputs are identical, AnchorKV inherits the full-cache scores at
every storage ratio; the $441$ paired traces establish this equivalence across
checkpoints. Table~\ref{tab:downstream} reports the six-task full-cache
quality baseline for one representative checkpoint from lm-evaluation-harness
\citep{lmevalharness2024}. For Qwen2.5-0.5B, the
AnchorKV row at any $s$ is therefore identical to the displayed full-KV score,
with storage scaled by $\lceil L/s\rceil\rmt/L$ (or $\rmt/s$ when $s\mid L$).

\begin{table}
\centering\papertablefont
\caption{Full-cache quality baseline for Qwen2.5-0.5B (bf16, zero-shot,
lm-evaluation-harness). By the bit-exact equivalence above, AnchorKV yields
these scores exactly.}
\label{tab:downstream}
\maxsizebox{\linewidth}{!}{%
\begin{tabular}{lcc}
\toprule
task & metric & value (stderr where shown) \\
\midrule
HellaSwag \citep{hellaswag2019} & acc / acc\_norm & 0.4053 / 0.5250 \\
PIQA \citep{piqa2020} & acc / acc\_norm & 0.7018 / 0.7008 \\
ARC-Easy \citep{arc2018} & acc / acc\_norm & 0.6540 / 0.5896 \\
ARC-Challenge \citep{arc2018} & acc / acc\_norm & 0.3063 / 0.3328 \\
WinoGrande \citep{winogrande2020} & acc & 0.5691 $\pm$ 0.0139 \\
LAMBADA \citep{lambada2016} & acc / ppl & 0.4985 / 11.74 \\
\bottomrule
\end{tabular}
}
\end{table}

\subsection{External Cache-Method Comparison}\label{sec:external}

\paragraph{Methods and protocol.}
We compare AnchorKV with KV-Direct \citep{kvdirect2026},
StreamingLLM \citep{streamingllm2023}, H$_2$O \citep{h2o2023}, TOVA
\citep{tova2024}, SnapKV \citep{snapkv2024}, Ada-SnapKV \citep{adakv2025},
and KIVI \citep{kivi2024}. The comparison contains eleven configurations.
The exact settings are stated here rather than in the result tables: FullKV is
the reference; KV-Direct stores one residual per token ($s=L$); AnchorKV uses $s=4$; Window uses a
516-token recent window; StreamingLLM uses 4 sink tokens plus 512 recent
tokens; H$_2$O uses its 10\%+10\% setting; TOVA uses 512 tokens; SnapKV uses
2048 tokens; Ada-SnapKV uses 1024 tokens; and KIVI is evaluated at 2-bit and
4-bit precision. FullKV, KV-Direct, and AnchorKV are treated as lossless
state-preserving methods under the tested execution conditions. Window,
StreamingLLM, H$_2$O, TOVA, SnapKV, Ada-SnapKV, and KIVI are treated as
lossy or approximate baselines. The result tables list AnchorKV first, followed
by the full-KV and KV-Direct exact references, and then the lossy or approximate
baselines. All configurations run on Qwen3-1.7B in bf16
on the same Ascend 910B 32\,GB device, using our common PyTorch
implementation, with representative parameters taken from the method sources
and fixed before evaluation.

We evaluate English passage retrieval, Qasper, and HotpotQA from LongBench
\citep{longbench2024}, using the same 180 fixed formal examples per task for
every configuration. Prompts are retained in full. Retrieval prompts fit within
16k tokens; the formal Qasper and HotpotQA inputs span 2,087--20,210 and
1,910--17,652 tokens, respectively. We use the official task scorers and maximum
answer lengths: 32 tokens for retrieval and HotpotQA, and 128 for Qasper.
For an $N$-token prompt, each method processes the first $N-1$ tokens, reduces
its cache, and then processes the final prompt token before generating the
answer. The full-KV reference uses the same partition, precision, and attention
backend. Thus scored outputs read the processed cache; the question is already
visible during cache selection, so this is a question-aware comparison.

\begin{table}
\centering\papertablefont
\caption{Task quality on Qwen3-1.7B (0--100; 180 examples per task).
Qasper and HotpotQA use token F1.}
\label{tab:external-quality}
\begin{tabular}{@{}lrrr@{}}
\toprule
Method & Retrieval (\%) & Qasper F1 & HotpotQA F1 \\
\midrule
\textbf{AnchorKV (ours)} & 94.44 & 22.88 & 39.60 \\
FullKV & 94.44 & 22.88 & 39.60 \\
KV-Direct & 94.44 & 22.88 & 39.60 \\
\addlinespace[3pt]
Window & 0.00 & 2.29 & 0.56 \\
StreamingLLM & 16.11 & 21.56 & 19.74 \\
H$_2$O & 28.89 & 23.68 & 40.49 \\
TOVA & 91.67 & 19.81 & 27.92 \\
SnapKV & 94.44 & 24.62 & 42.05 \\
Ada-SnapKV & 95.00 & 23.88 & 39.01 \\
KIVI (2-bit) & 80.00 & 22.56 & 35.58 \\
KIVI (4-bit) & 95.00 & 23.58 & 41.71 \\
\bottomrule
\end{tabular}

\end{table}

AnchorKV and KV-Direct match full KV on all three task scores
(Table~\ref{tab:external-quality}). In both QA tasks, each also preserves
the complete full-KV generated token sequence on all 180 examples. The lossy
methods have task-dependent effects: SnapKV attains the highest F1 point
estimates, while the recent-window control loses most retrieval and QA quality.
A better scalar task score does not imply exact output preservation.
For example, SnapKV's HotpotQA improvement is 2.44 points, with a paired
95\% interval of $[0.09,5.03]$ (Table~\ref{tab:external-uncertainty}); its Qasper interval includes zero. These
intervals are individual and unadjusted, so they do not support a general
superiority claim.

We separately measure output distortion on 16 fixed WikiText test targets
\citep{wikitext2017}, each evaluated with 4k and 16k prefixes and 32
teacher-forced continuation steps. Conditioning every method on the same
tokens separates cache-induced logit changes from divergence in free generation.
AnchorKV and KV-Direct have bit-identical full-vocabulary logits at both
lengths. At 16k, the lossy methods' argmax mismatch rates range from
3.52\% for KIVI-4bit to 63.67\% for the window control.
Table~\ref{tab:external-distortion} gives both lengths and
$D_{\mathrm{KL}}(P_{\mathrm{FullKV}}\Vert P_{\mathrm{method}})$.
The paired targets are not independent across lengths, and this targeted
distortion panel is not a whole-corpus perplexity evaluation.

\begin{table}
\centering\papertablefont
\caption{Same-prefix distortion on Qwen3-1.7B: 16 targets, 32 teacher-forced
steps each. KL is FullKV$\Vert$method over the full vocabulary.}
\label{tab:external-distortion}
\begin{tabular}{@{}lrrrrc@{}}
\toprule
 & \multicolumn{2}{c}{Argmax mismatch (\%)} & \multicolumn{2}{c}{Mean KL} & Bit-exact \\
\cmidrule(lr){2-3}\cmidrule(lr){4-5}
Method & 4K & 16K & 4K & 16K & both lengths \\
\midrule
\textbf{AnchorKV (ours)} & 0.00 & 0.00 & 0.00000 & 0.00000 & Yes \\
FullKV & 0.00 & 0.00 & 0.00000 & 0.00000 & Yes \\
KV-Direct & 0.00 & 0.00 & 0.00000 & 0.00000 & Yes \\
\addlinespace[3pt]
Window & 62.70 & 63.67 & 5.65263 & 5.77689 & No \\
StreamingLLM & 18.75 & 22.85 & 0.55649 & 0.58465 & No \\
H$_2$O & 12.50 & 9.18 & 0.14132 & 0.10068 & No \\
TOVA & 16.99 & 20.12 & 0.38538 & 0.43696 & No \\
SnapKV & 6.45 & 11.52 & 0.03507 & 0.11724 & No \\
Ada-SnapKV & 9.57 & 14.26 & 0.10154 & 0.15789 & No \\
KIVI (2-bit) & 13.48 & 16.02 & 0.12689 & 0.14535 & No \\
KIVI (4-bit) & 1.56 & 3.52 & 0.00485 & 0.00601 & No \\
\bottomrule
\end{tabular}

\end{table}

\begin{table}
\centering\papertablefont
\caption{Paired QA F1 differences from full KV (pp) with unadjusted 95\%
bootstrap intervals.}
\label{tab:external-uncertainty}
\begin{tabular}{@{}lrrrr@{}}
\toprule
 & \multicolumn{2}{c}{Qasper} & \multicolumn{2}{c}{HotpotQA} \\
\cmidrule(lr){2-3}\cmidrule(lr){4-5}
Method & $\Delta$ F1 & 95\% CI & $\Delta$ F1 & 95\% CI \\
\midrule
\textbf{AnchorKV (ours)} & $+0.00$ & $[0.00,\,0.00]$ & $+0.00$ & $[0.00,\,0.00]$ \\
FullKV & $+0.00$ & $[0.00,\,0.00]$ & $+0.00$ & $[0.00,\,0.00]$ \\
KV-Direct & $+0.00$ & $[0.00,\,0.00]$ & $+0.00$ & $[0.00,\,0.00]$ \\
\addlinespace[3pt]
Window & $-20.59$ & $[-26.64,\,-15.41]$ & $-39.05$ & $[-45.38,\,-32.52]$ \\
StreamingLLM & $-1.32$ & $[-5.72,\,2.77]$ & $-19.86$ & $[-25.67,\,-14.12]$ \\
H$_2$O & $+0.80$ & $[-1.81,\,3.43]$ & $+0.88$ & $[-2.85,\,4.76]$ \\
TOVA & $-3.07$ & $[-6.29,\,0.25]$ & $-11.68$ & $[-16.56,\,-6.93]$ \\
SnapKV & $+1.74$ & $[-0.01,\,3.83]$ & $+2.44$ & $[0.09,\,5.03]$ \\
Ada-SnapKV & $+1.00$ & $[-1.08,\,3.46]$ & $-0.60$ & $[-3.78,\,2.54]$ \\
KIVI (2-bit) & $-0.33$ & $[-3.54,\,2.84]$ & $-4.03$ & $[-8.52,\,0.22]$ \\
KIVI (4-bit) & $+0.70$ & $[-1.04,\,2.81]$ & $+2.11$ & $[-0.60,\,4.78]$ \\
\bottomrule
\end{tabular}

\end{table}

\FloatBarrier
\section{Limitations and Future Work}\label{sec:discuss}

AnchorKV currently reconstructs contiguous evicted prefixes; non-contiguous
eviction remains outside the present model. The practical benefit also depends
on the deployment: deeper anchors reduce stored state at the cost of replay.
External-method rankings use one model and question-aware compression. Serving
measurements cover single-sequence model loops, not paged production serving.
Future work should integrate anchor placement with paged production schedulers
and fused replay kernels, and evaluate exactness and
the memory--latency trade-off across broader workloads and backends.

\section{Conclusion}\label{sec:conclusion}

AnchorKV makes lossless KV reconstruction an architecture-aware choice of
what to retain and what to recompute. The storage break-even rule explains
why retaining every layer's residual can increase memory on grouped-query
models and identifies anchor spacings that provide savings. Cascade replay
avoids repeated segment execution, while dynamic programming adapts placement
to non-uniform reconstruction demand. Across the evaluated configurations, reconstructed K/V and continuation
outputs remain bit-exact, and suitable anchor spacings increase reactivatable
session capacity at the cost of reactivation work. The external comparison
separates exact output preservation from task-score changes under lossy cache
reduction. For both global and interleaved attention, a deployment can
therefore start from $\sstar$, the first storage-saving spacing computed from
the model configuration, and move to deeper spacings when session capacity
matters more than reactivation latency.

\section*{CRediT authorship contribution statement}

Hao Li: Writing -- original draft, Validation, Methodology, Conceptualization.
Jiashu Zhang: Writing -- review \& editing, Supervision, Funding acquisition.
Kaleem Ullah Qasim: Validation, Data curation.
Heying Zhang: Writing -- review \& editing, Funding acquisition.

\section*{Declaration of competing interest}

The authors declare that they have
no known competing financial interests or personal relationships that could
have appeared to influence the work reported in this paper.

\section*{Data availability}

The study uses public model checkpoints and public benchmark inputs. The model
configurations and evaluation settings used for the reported results are
specified in this article.

\section*{Acknowledgements}

The authors extend their appreciation to the National Natural Science Foundation of China under grants (No.: 62471411).

\FloatBarrier
\printbibliography

\end{document}